\documentclass[fleqn,usenatbib]{mnras}

\usepackage{newtxtext,newtxmath}

\usepackage[T1]{fontenc}
\usepackage{ae,aecompl}
\usepackage{xcolor}

\usepackage{graphicx}
\usepackage{amsmath}	

\usepackage{multicol}
\usepackage{graphicx}
\usepackage{subcaption}
\usepackage{pdflscape}
\usepackage{threeparttable, tablefootnote}
\title[MMA and Galactic RSGs]{Red Supergiant Candidates for Multimessenger Monitoring of the Next Galactic Supernova }

\author[Sarah Healy]{Sarah Healy$^{1}$\thanks{E-mail:healys@vt.edu},
    Shunsaku Horiuchi$^{1,2}$,
    Marta Colomer Molla$^{3}$,
    Dan Milisavljevic$^{4,5}$,
    Jeff Tseng$^{6}$,
    \newauthor
    Faith Bergin$^{5}$,
    Kathryn Weil$^{5}$,
    Masaomi Tanaka$^{7}$,
    and Sebastián Otero$^{8}$
\\
$^{1}$Center for Neutrino Physics, Department of Physics, Virginia Tech, Blacksburg, VA 24061, USA \\
${^2}$Kavli IPMU (WPI), UTIAS, The University of Tokyo, Kashiwa, Chiba 277-8583, Japan
$^{3}$Université Libre de Bruxelles, 1050 Bruxelles, Belgium\\
$^{4}$Department of Physics and Astronomy, Purdue University, West Lafayette, IN 47907, USA \\
$^{5}$Integrative Data Science Initiative, Purdue University, West Lafayette, IN 47907, USA\\
$^{6}$Department of Physics, Oxford University, Oxford OX1 3RH, UK\\
$^{7}$Astronomical Institute, Tohoku University, Sendai 980-8578, Japan\\
$^{8}$American Association of Variable Star Observers (AAVSO), Cambridge, MA 02138, US}

\date{Accepted 2024 March 7. Received 2024 March 6; in original form 2023 July 28}

\pubyear{2023}
\begin{document}
\label{firstpage}
\pagerange{\pageref{firstpage}--\pageref{lastpage}}
\maketitle

\begin{abstract}
We compile a catalog of 578 highly probable and 62 likely red supergiants (RSGs) of the Milky Way, which represents the largest list of Galactic RSG candidates designed for continuous follow-up efforts
to date. We match distances measured by Gaia DR3, 2MASS photometry, and a 3D Galactic dust map to obtain luminous bright late-type stars. Determining the stars' bolometric luminosities and effective temperatures, we compare to Geneva stellar evolution tracks to determine likely RSG candidates, and quantify contamination using a catalog of Galactic AGB in the same luminosity-temperature space. We add details for common or interesting characteristics of RSG, such as multi-star system membership, variability, and classification as a runaway. As potential future core-collapse supernova progenitors, we study the ability of the catalog to inform the Supernova Early Warning System (SNEWS) coincidence network made to automate pointing, and show that for 3D position estimates made possible by neutrinos, the number of progenitor candidates can be significantly reduced, improving our ability to observe the progenitor pre-explosion and the early phases of core-collapse supernovae.

\end{abstract}


\section{Introduction}

Red Supergiants (RSGs) are one of the last evolutionary stages of massive stars, and have been directly linked to Type-IIP/L supernovae (SNe) through matching SN locations with archival pre-explosion images \citep{Smartt_2004,Smartt_2009, Smartt_2015,daviesandbeasor_2020,beasor_2018,Li_2007,vandyk_2017notreview,van_2012}; a comprehensive summary of can be found in \cite{Van_Dyk_2017,Smartt_2015}. The connection between RSGs and core-collapse supernovae (CCSNe) links two stellar phases between a hugely disruptive explosion, which enables studies of stellar evolution as well as the mechanisms that cause CCSN. Furthermore, as progenitors of SNe, RSG properties impact the formation of stellar mass black holes and neutron stars, themselves sources of gravitational waves through possible eventual mergers. 

The next Galactic CCSN will represent a once-in-a-generation opportunity to study important progenitor-SN connections in unprecedented detail. 
Such CCSNe are rare, but they allow us to study neutrinos \citep[e.g.][]{Scholberg_2012}, gravitational waves \citep[GW; e.g.][]{Ott_2009,Kotake_2013}, and nuclear gamma rays \citep[e.g.][]{Gehrels_1987,Horiuchi_2010}. These observables allow us to probe far more completely inside the stellar photosphere. The neutrinos are key since they are emitted from the proto-neutron star deep within the progenitor. They are also emitted before the SN starts and the neutrinos can be used to point to the location in the sky with an error circle of a few to ten degrees \citep{Beacom_1999,Bueno_2003,Tomas_2003,Mukhopadhyay_2020}. These make neutrinos ideal as an early warning trigger for the SN and is the backbone of multi-messenger observation strategies for CCSNe. In addition, the final stage of Si burning in massive stars emits neutrinos with $\sim 5 \times 10^{50}$ erg \citep{Arnett_1989}, which can be detected for progenitors within a few kpcs \citep{Odrzywolek_2004,Kharusi_2021,KamLAND:2015dbn}, acting as another layer of early trigger. Various aspects, including predictions, implementations, and detectability of the multi-messenger signals of Galactic and nearby CCSNe have been explored over the years \citep{Adams_2013,Nakamura_2016,Kharusi_2021}. 

To most effectively capture this opportunity, awareness of the properties and spatial distributions of RSGs is crucial. With such knowledge, the rapid identification of the SN progenitor would become more realistic, e.g., mirroring the strategies in the gravitational wave community with lists of galaxies for follow-up searches (\citealt{2011CQGra..28h5016W}; and more recently \citealt{GLADE_2018}). The effective use of a pre-compiled target list would allow, among others, monitoring of the earliest light curves of the progenitor/SN, which are crucial for accurate reconstruction of the CCSN evolution and determining progenitor properties \citep[e.g.,][]{Tominaga_2011}. 

However, our knowledge of RSGs remains incomplete. In particular, the spatial and luminosity distributions are not well known. Spectra from massive stars are limited, and even more so are those with assigned spectral types. While the temperature scale of Galactic RSGs 
has been a topic of study for decades \citep{johnson_1964,johnson_1966,flower_1975,flower_1977,lee_1970}, and has been updated in a series of recent papers---\cite{Levesque_2005_17_ref_15} and \cite{Levesque_2006_18} compared the strength of the TiO bands in the optical to that of MARCS’s model atmospheres---the debate about the precise temperatures of Galactic RSGs is still ongoing \citep{taniguchi_2021_a,taniguchi_2021_b}. It is also known that the most evolved RSGs have substantially higher levels of circumstellar extinction, but the dust sizes and distributions are not well modeled. Since dust estimations directly affect luminosity calculations, this can lead to errors in a star's bolometric luminosity, mass, and stellar radius \citep{Beasor_2016_7}. 

While the extensive information that can be obtained from Galactic CCSNe cannot be matched by extragalactic ones, it is worth noting that RSG samples in the Magellanic clouds, M31, and M33 are estimated to be near-complete. The number of known RSGs in the LMC and SMC was initially a few tens \citep{feast_1980,catchpole_1981,wood_1983,pierce_2000} and increased to a couple hundred \citep{massey_2002,massey_olsen_2003,yang_jiang_2011,yang_jiang_2012,neugent_2012,gonzalez-fernandez_2015} but the more recent work of \cite{yang_2019,yang_2021} identified an approximately 90 percent complete sample with 1405 (SMC) and 2974 (LMC). Similarly, work done in M31 and M33 initially selected 437 (M31) and 776 (M33) candidates \citep{massey_2006,massey_2007,massey_2009,drought_2012} with 255 of those confirmed to be RSGs by \cite{massey_evans_2016}. However, \cite{ren_2019} found a lack of known faint stars motivating the identification of a near-complete sample containing 5498 (M31) and 3055 (M33) RSGs \citep{Ren_2021}.

Mapping the Milky Way’s RSGs has been attempted in the past. For example, those compiled with a greater focus on pre-explosion studies, including \citealt{Messineo_2019} with 889 late-type bright stars of which $\sim$382 are RSG candidates, and \cite{messineo_2023} with 203 bright late-type stars of which 20 are candidate RSGs. Also, lists exist that are focused more on post-explosion studies, including \citealt{Nakamura_2016}'s 212 stars and \citealt{Mukhopadhyay_2020}'s 31 stars. We build on the methods of \citealt{Messineo_2019} but 
design our methodology to minimize losing a true RSG even at the expense of keeping more non-RSG contaminates. To further augment our list, we include a second method for search. We also compile or estimate additional stellar characteristics, e.g., variability, radius, multi-star system, cluster membership, runaway status, presence of a significant magnetic field, mass-loss rate, and classification as a runaway star. We present a systematic list of Galactic binaries. Finally, we look into our final samples’ angular separation compared to the angular resolution anticipated from future CCSN neutrino detections in preparation for use in multi-messenger astronomy. Further discussion of Galactic RSG surveys in the literature and comparisons to our work is in Section \ref{catalog comparison}.

A comprehensive list of evolved massive stars in the Milky Way galaxy will have numerous broader benefits beyond multi-messenger astronomy. For example, it can be used to study the effects of massive star evolution, binary influence, as well as metallicity. RSGs trace stars with masses from about 9 to 40 $\rm M{\textsubscript{\(\odot\)}}$, i.e., ages of 4 to 30 Myr \citep{Ekstrom_2012,Chieffi_2013}. The binary frequency of RSGs can be used to test interaction rates when compared to unevolved massive stars. At the point of writing this paper, approximately a dozen Galactic RSG binaries are confirmed \citep{Neugent_2019}. The main factor in making these improvements is having a relatively complete Galactic RSGs population that also spans various metallicities. 

This paper is organized as follows. In Section \ref{2:methods}, we compile our list of bright and cool stars and measurements of their distance, photometry, and dust extinction. We determine our most confident sample using luminosity, stellar evolution tracks, and note stars with typical RSG characteristics in Section \ref{3:results}. In Section \ref{discussion and interperation}, we discuss specific objects of interest, the sample's spatial distribution, the range of estimated radii, and mass loss. The implications for the list and coordinated use with multi-messenger astronomy (MMA) for Galactic CCSNe and massive star astronomy are discussed in Section \ref{catalog comparison} \ref{MMA}. We conclude in Section \ref{summary}.

\section{Methods} \label{2:methods}
We aim to determine whether a star is a RSG by using its effective temperature and bolometric luminosity. We first collect a sample of luminous late-type stars, which we define by spectral type K or M and luminosity class I. Based on spectral types, $\rm T_{eff}$ is estimated with the temperature scale for RSGs determined in \cite{Levesque_2005_17_ref_15}. From there, we collect photometry measurements, model dust extinction, and estimate the intrinsic bolometric magnitude and $\rm T_{eff}$, which are crucial to determining whether the star is more likely an RSG or contamination from a non-RSG.

\subsection{Data Selection} 
We consider two data selection methods, each designed to create a starting list of stars with RSG-like properties. The first ``compilation-base'' method focuses on selecting stars with predetermined spectral types matching the expectations of RSGs, i.e., a collection of K-M types stars with luminosity class I. However, spectral classifications of Galactic massive stars are relatively small in size. To remove this limitation, we use a second ``Gaia-based'' method, which focuses on obtaining stars whose position on the HR (or pseudo HR) diagram matches the expectations of RSGs. To this end, we utilize Gaia photometry and color to place a cut on Gaia DR3 stars to obtain stars that compare to K-M stars with luminosity class I. We describe in detail the two methods below and summarize total counts in Table \ref{table:sample-reduction}.

\subsubsection{Method 1: compilation-based sample}

We begin by collecting Galactic spectral catalogs that use the Morgan–Keenan (MK) classification system \citep{morgan_1943}. This includes those who took spectra in the Near-infrared \citep{figer_2006_ref_46,messineo_2014_ref_18}, infrared \citep{Blum_2003_ref_1,verhoelst_2009_ref_16}, K-band specific \citep{Liermann_2009_ref_6,negueruela_2010_ref_7}, and optical \citep{Levesque_2005_17_ref_15} and with low \citep{Clark_2005_ref_2,messineo_2008_ref_19} to high \citep{de_burgos_2020_ref_21,negueruela_2011_ref_8,alonso_santiago_2017_ref_23} resolution. Some look specifically at Galactic clusters including Westerlund I \citep{westerlund_1987_ref_41,Clark_2005_ref_2,mengel_2007_ref_20}, scutum-Cruz arm \citep{Clark_2009_ref_3}, RSGC1-3 \citep{negueruela_2010_ref_7,Davies_2008_ref_25}, while others focused generally on stars in OB associations \citep{massey_2001_ref_17,garmany_1992_ref_35}. We set no requirement of what bands were used as indicators; some use the TiO band \citep{Elias_1985_ref_14,white_1978_ref_38}, the Ca II triplet \citep{Dorda_2018_ref_5,Dorda_2016_ref_4,negueruela_2012_ref_9}, the CO band \citep{negueruela_2012_ref_9,Davies_2007_ref_24,Davies_2008_ref_25}, or some combination of bands \citep{kleinmann_1986_ref_11,rayner_2009_ref_10,massey_2001_ref_17}. 

In an effort to be as complete as possible, we also included catalogs that were compilations of others \citep{humphreys_1978_ref_13,jura_1990_ref_12}. \cite{skiff_2014} is the largest such catalog, containing 1,058,791 entries, and not all entries can be found in Simbad\footnote{These stars are still retained in our analysis.}. Entries in the catalog have been converted to the MK type where possible.

\begin{table}
\begin{threeparttable}
\centering
\caption{Progression of sample size moving through steps building up to calculations of bolometric luminosity.}\label{table:sample-reduction}
\begin{tabular}{l|c|c} 
      \hline
    Cuts&Compilation Based&Gaia Based\tnote{1}\\
    \hline
    Bright Late-type Stars&2,051& $\rm \sim2.2\times10^7$\\
    K or M Spectral Type\tnote{2}&2,051&3,499\\
    Matches in Gaia&2,046&3,499\\
    Distance Cut&1,480&3,425\\
    Cross Matches in 2MASS&1,421&3,410\\
    Final RSGs\tnote{3} &555&38\\
    \hline
\end{tabular}
\begin{tablenotes}\footnotesize
 \item [1] Includes contributions from DR2 and DR3
 \item [2] This step only applies to the Gaia based method
 \item [3] Includes 15 duplicates between the two methods
 \end{tablenotes}
\end{threeparttable}
\end{table}

We restrict ourselves to K and M type stars with at least one classification of or including luminosity class I. We do not restrict the inclusion of spectra based on location in the sky. Special care was made to ensure that if there were repeats in original source classifications between these catalogs or with the earlier list, only one entry was included. A portion of the list can be found in Table \ref{tab:sampleone}. 

Not all observations were unique and a single star could have a dozen classifications, so we implemented a method for adopting spectral types. For catalogs with well-studied objects, we adopted spectral types based on the classification assigned in those catalogs. We gave preference to \cite{Levesque_2005_17_ref_15} as they used both moderate-resolution optical spectrophotometry and the MARCS stellar atmosphere models fits to determine the spectral type. Thereafter, we prioritize the catalogs with well-studied classifications in \cite{Dorda_2016}, \cite{Dorda_2018_ref_5}, \cite{jura_1990_ref_12}, \cite{Elias_1985_ref_14}, and \cite{humphreys_1978_ref_13}. 

If none of the previous works provide spectral types, we adopt a mean spectral type weighted by uncertainties. This keeps in mind that classifications derived from fewer than the 8 TiO bands or from bands that have less characteristic behavior for massive stars and those who have observational limitations, like \cite{keenan_1989_ref_26}, generally have larger uncertainties between $\pm$ 1-2 subtypes (usually $\pm$ 0.5 subtypes for classification done by visual inspection of the whole 8 TiO band spectrum). We include \cite{white_1978_ref_38}, who, along with stating previous spectral classification, finds the mean spectral type from the 8-color indices, which use a photoelectric system defined by eight narrow bands between 0.7 and 1.1 $\mu$m to determine two-dimensional spectral classification. We do not adopt 2-D spectral types but use them to inform mean values for entries with many classifications. This is accomplished by converting K0-M10 to a numerical scale, calculating the weighted average, and then rounding to the nearest half-integer, except for special cases with 4 or more entries where a single spectral type appears a majority of the time. 

Initial celestial positions were taken directly from the source paper, but after combining all catalogs, some variance in RA and Dec was observed, partially due to the different epochs used. In order to improve the positions, we matched the stars listed in Table \ref{tab:sampleone} to Simbad based on the alias of the star and took the corresponding position with epoch J2000.0, also shown in Table \ref{tab:sampleone}. 

In summary, our compilation-based Method provided 2,051 stars with either K or M spectral type and at least one classification of luminosity class I.

\begin{table*}
\begin{threeparttable}
 \caption{Summary of late-type bright stars compiled from both method whose distance uncertainty meet the requirements detailed in \S\ref{2.2}. For stars from the compilation-based method, we list all spectral types recorded along with the mean adopted spectral type. Spectral types from \citep{houk_1978} for stars in the Gaia-based sample are included with the adopted spectral type set equal to the spectral type.  Alias, Gaia DR3 IDs, RA, Dec, parallax, and distance are included along with the $\rm T_{eff}$ determined using the adopted spectral type. The full table is publicly available at \url{https://github.com/SNEWS2/candidate_list}.}
  \label{tab:sampleone}
 \begin{tabular}{llllllllll}
  \hline
  Alias&Gaia DR3 ID&RA&Dec&Parallax&SpType\tnote{1}&SpType\_a\tnote{2}&Source\tnote{3}&{\tt r\_med\_geo}\tnote{4}&$\rm T_{eff}$\\
  \hline
   IRAS 20315+4026 & 
    2064747910568516608 & 308.35 & 40.61 & 0.18 & M0: & M0 & 0 & 3809.8 & 3790.0 \\
    * psi01 Aur & 970675154497903616 & 96.22 & 
    49.29 & 0.44 & M0, K4, K5, ... & M0 & 0, 36, 38, ... & 2021.5 & 3790.0 \\
    *  6 Gem & 3425055656275589632 & 93.08 & 
    22.91 & 0.56 & 
    M2, M1, M2+, ... & M0 & 0, 36,
    38, ... & 1757.05 & 3790.0 \\
    DO 42870 & 2010457973461961344 & 349.63 & 
    58.55 & 0.26 & M1, M1 & M1 & 0, 5 & 3393.6182 & 3745.0 \\
    '[NBM54]' 44 & 4097664886515067392 & 273.65 & 
    -16.36 & 0.13 & M1 & M1 & 0 & 5085.879 & 3745.0 \\
    V* NR Vul & 2020687421645374720 & 297.55 & 
    24.92 & 0.32 & M2, M1, M1, ... & K3 & 0, 36, 38, ... & 2774.4 & 3985.83 \\
    DO 42870 & 2010457973461961344 & 349.63 & 
    58.55 & 0.26 & M1, M1 & M1 & 0, 5 & 3393.62 & 3745.0\\
    CD-30  5114 & 5598579347402547072 & 117.05 & 
    -30.83 & 0.28 & K2.5 & K2.5 & 0 & 3267.55 & 4015.0 \\
    HD  87438 & 5255870712734219648 & 150.71 & 
    -62.16 & 0.85 & K3 & K3 & 0 & 1175.17 & 3985.83 \\
    HD 119796 & 5865517646532509568 & 206.8 & 
    -62.59 & 0.25 & K0, K0, K0 & K0 & 0, 26 & 3600.67 & 4185.0 \\
  \hline
 \end{tabular}
 \begin{tablenotes}\footnotesize
 \item [1] List of all spectral types matched to an object and the corresponding sources are in the same order under the Source column
 \item [2] Adopted spectral type based on the mean of SpType
 \item [3] 1 \protect\cite{Blum_2003_ref_1}, 2 \protect\cite{Clark_2005_ref_2}, 3 \protect\cite{Clark_2009_ref_3}, 4 \protect\cite{Dorda_2016_ref_4}, 5 \protect\cite{Dorda_2018_ref_5}, 6 \protect\cite{Liermann_2009_ref_6}, 7 \protect\cite{negueruela_2010_ref_7}, 8 \protect\cite{negueruela_2011_ref_8}, 9 \protect\cite{negueruela_2012_ref_9}, 10 \protect\cite{rayner_2009_ref_10}, 11 \protect\cite{kleinmann_1986_ref_11},12 \protect\cite{jura_1990_ref_12}, 13 \protect\cite{humphreys_1978_ref_13}, 14 \protect\cite{Elias_1985_ref_14}, 15 \protect\cite{Levesque_2005_17_ref_15}, 16 \protect\cite{verhoelst_2009_ref_16}, 17 \protect\cite{massey_2001_ref_17}, 18 \protect\cite{messineo_2014_ref_18}, 19 \protect\cite{messineo_2008_ref_19}, 20 \protect\cite{mengel_2007_ref_20}, 21 \protect\cite{de_burgos_2020_ref_21}, 22 \protect\cite{beauchamp_1994_ref_22}, 23 \protect\cite{alonso_santiago_2017_ref_23},  24 \protect\cite{Davies_2007_ref_24}, 25 \protect\cite{Davies_2008_ref_25}, 26 \protect\cite{keenan_1989_ref_26}, 27 \protect\cite{alexander_2009_ref_27}, 28 \protect\cite{bidelman_1957_II_ref_28}, 29 \protect\cite{Bidelman_1957_I_ref_29}, 30 \protect\cite{halliday_1955_ref_30}, 31 \protect\cite{medhi_2007_ref_31}, 32 \protect\cite{ginestet_1999_ref_32}, 33 \protect\cite{ginestet_1997_ref_33}, 34 \protect\cite{fawley_1974_ref_34}, 35 \protect\cite{garmany_1992_ref_35}, 36 \protect\cite{humphreys_1970a_ref_36}, 37 \protect\cite{humphreys_1972_ref_37}, 38 \protect\cite{white_1978_ref_38}, 39 
 \protect\cite{marco_2013_ref_39}, 40 \protect\cite{boulon_1963_ref_40}, 41 \protect\cite{westerlund_1987_ref_41}, 42 \protect\cite{marco_2014_ref_42}, 43 \protect\cite{Currie_2010_ref_43}, 44 \protect\cite{gonzalez-fernandez_2012_ref_44}, 45 \protect\cite{humphreys_1970b_ref_45}, 46 \protect\cite{figer_2006_ref_46}
 \item [4] {\protect{\tt r\_med\_geo}: geometric and photometric distance estimate from Gaia DR3's parallaxes \protect\citep{bailer_jones_2021}}
 \end{tablenotes}
 \end{threeparttable}
\end{table*}

\begin{figure}
    \makebox[\textwidth][l]{\includegraphics[width=0.5\textwidth]{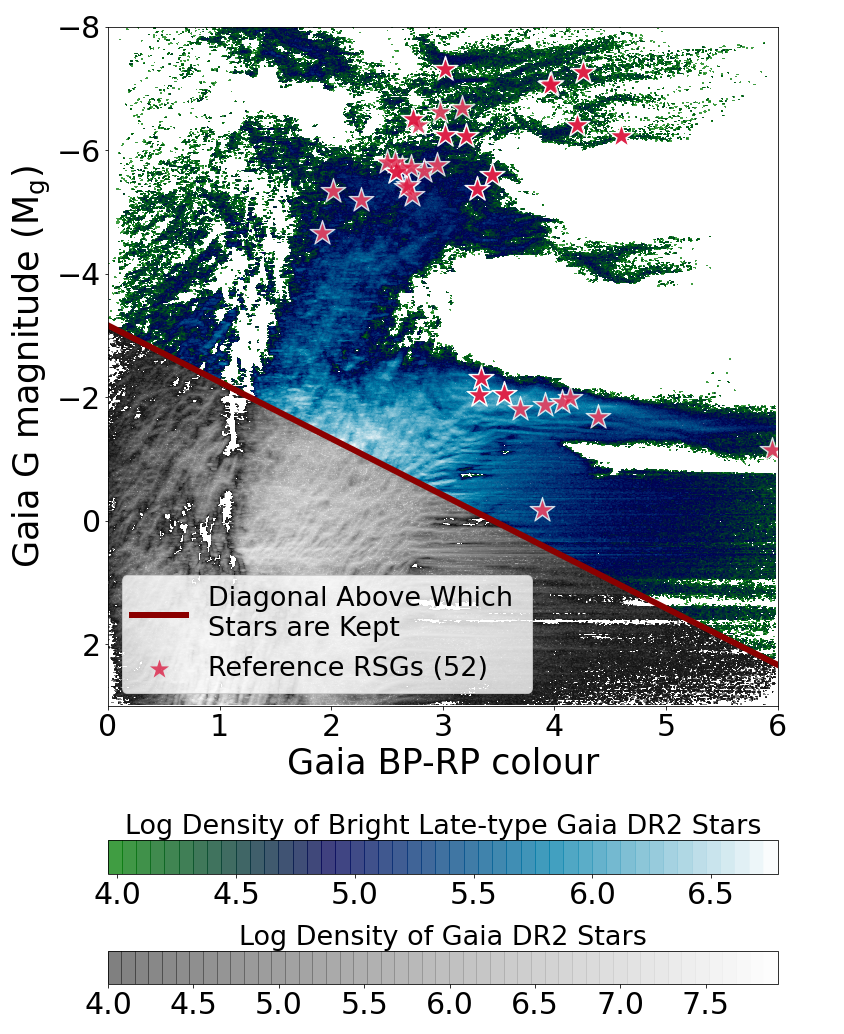}}
    \caption{ Log of the density distribution of Gaia DR2 stars in the colour-magnitude diagram of absolute Gaia G passband magnitudes (near-ultraviolet to near-infrared) and BP-RP colour. Overlaid are 52 stars in our reference RSG sample (see selection criteria in \S\ref{gaiabased}), plotted as red stars. Our selection cut is indicated by the solid thick dark red line (see Eq.~(\ref{eq:cut})). Stars above the cut in the green-blue shade are retained as they have temperatures and luminosity similar to those in our reference RSG sample. 
    \label{fig: Luminosity Cuts to Gaia}}
\end{figure}

\subsubsection{Method 2: Gaia-based sample} \label{gaiabased}

In an effort to expand our list further, we utilized the vast stellar database of Gaia (DR2/DR3) and a subset of our compilation-based sample to select RSG candidates. We use Gaia's G passband (roughly 330-1050 nm) magnitude and BP-RP (330-680 nm; 630-1050 nm) colour band for rough estimates of luminosity and effective temperature, respectively. While the absolute G magnitude does not provide total bolometric luminosity, we use it to indicate where a star would be if we had included all magnitudes from other bands; the Gaia G band works well for this purpose as it spans near-ultraviolet to near-infrared. Larger differences between passbands correlate with cooler stellar temperatures. The star’s temperature can be inferred as the larger the difference between passbands, the cooler the star. These are shown in Fig.~\ref{fig: Luminosity Cuts to Gaia}. Here, the color indicates the log of stellar density. 

We used a subset from our compilation-based sample to identify the region in the colour-magnitude diagram (CMD) corresponding to late-type bright stars. For this purpose, we used stars whose RSG status is more confident, either low uncertainty or how well-analyzed it is. This includes four Galactic clusters: RSGC1, RSGC2 (Stephenson 2), RSGC3, and NGC 7419, which cover the range of expected luminosity as well as include masses $\rm \simeq 9 M{\textsubscript{\(\odot\)}}$ to $\rm \lesssim 25 M{\textsubscript{\(\odot\)}}$. These clusters are some of the most well-studied in terms of RSG population; most stars included in these regions have follow-up spectra, allowing us to restrict our set further to those that are or have a follow-up observation. Ultimately we use 156 stars compiled from \cite{Levesque_2005_17_ref_15}, \cite{humphreys_1978_ref_13}, \cite{figer_2006_ref_46} (RSGC1), \cite{negueruela_2012_ref_9} (RSGC2), \cite{Clark_2009_ref_3} (RSGC3), and \cite{marco_2013_ref_39} (NGC 7419). We call this our reference RSG sample. 

The reference RSGs were matched to Gaia DR3 (DR2) objects with both G magnitude and estimated extinctions, which reduced the size to 30 (52) stars. The DR2 matches are shown by red stars in Fig.~\ref{fig: Luminosity Cuts to Gaia} and reveal the region where RSGs populate. The strong bimodality of the RSGs on the CMD presented in Fig.~\ref{fig: Luminosity Cuts to Gaia} likely results from the nature of stellar metallicity distribution and its effects on stellar evolution. Studies of the metallicity of the horizontal branches of globular clusters (using Gaia DR2 data; \citealt{gaia_col_2018c} and \citealt{zinn_1985}) and halo stars (using Gaia DR2 data; \citealt{gaia_col_2018c} and TGAS data with RAVE and APOGEE; \citealt{bonaca_2017}) show the distribution is double-peaked. It is also known that the evolutionary tracks of stars are highly dependent on metallicity \citep{el_eid_2009,langer_2012,maeder_2009}. The double-peaked metallicity distribution of stars before the RSG phase and the impact metallicity has on stellar evolution suggests the distribution of stars in the CMD is similarly bimodal as seen in Fig.~\ref{fig: Luminosity Cuts to Gaia}.

Including both regions, we derive a cut to select Gaia stars in the CMD that are likely to be RSGs. While different cuts can be found for different samples, we choose to use the cut derived from DR2 for both Gaia datasets as few of our reference RSGs have estimated extinction in DR3. This limitation is primarily driven by the nature of bright stars and the high uncertainties they would likely have when methods for determining extinction values are based on training sets that exclude massive stars, as is the case for Gaia DR2 and DR3 \citep{gaia_dust_2022,gaia_extinction_2013,gaia_extinction_updates_2022}. As our cut determined from Gaia DR2 is derived from absolute G mag and BP-RP colours, it is applicable to any dataset containing those measurements. It includes all in the DR3-matched reference set. Our data cut is: 
\begin{equation}\label{eq:cut}
    M_G \le 0.916(BP-RP) + (-3.165)
\end{equation}
and shown by a thick solid line in Fig.~\ref{fig: Luminosity Cuts to Gaia}. Stars above the cut are bright and cool enough to be candidates RSGs. Figure \ref{fig: Luminosity Cuts to Gaia} shows how all our reference RSGs are above the cut. All stars are retained regardless of uncertainty; we discuss the accuracy of distance in \S\ref{2.2}. The cut makes a homogeneous sample in terms of extinction and magnitude. Still, those values are not used for later analysis, as our goal is to determine bright late-type stars homogeneously. The datasets were merged, preserving only unique values. 

Our cut in the CMD gives a degree of confidence in the luminosities of selected stars. Thus, we only need to take stars with appropriate spectral types. However, our CMD provides only rough estimates for spectral types. Thus, we obtained spectral types from the Henry Draper (HD) catalogs. Over multiple publications, the HD catalogs (1918-1924), Henry Draper Extension (HDE; 1925-1936), and Henry Draper Extension Charts (HDEC; 1934) determined spectral classification and rough positions for 272,150 stars. These spectral classifications are based on the Harvard system, which has slightly different notations for spectral types than the MK system. For consistency, we use the Michigan Catalog of HD (MCHD) (1975-1999), which contains $\rm \sim$161,000 HD stars that were reclassified to the MK system using the spectra from HD, HDE, and HDEC \citep{houk_1978}. 

Matching was done for all stars to available M and K spectral type objects from MCHD providing an additional, though not entirely unique from the compilation-based method, set of 3,499 candidates. The overlap with the compilation-based method is $<$100 stars, reinforcing that our Gaia-based method provides us with valuable new candidate RSGs. 

\subsection{Distance related cuts with Gaia} \label{2.2}
We utilized the Gaia DR3 catalog, which contains over 1.8 billion sources \citep{gaia_mission_2016,gaia_dr3_summary_2022} to provide reliable parallaxes for both the compilation-based and Gaia-based samples, including uncertainties for each star. 
After converting coordinates from J2000 to J2016, Gaia matches were found within a 2" radius of each late-type star, producing objects for $\rm \sim 96\%$ of the compilation-based method and, by design, 100\% of the Gaia-based method.

The Renormalised Unit Weight Error (RUWE), described in the technical note \citep{Lindegren_2018}, is recommended by the Gaia team as a proxy for a good astrometric solution to a star. The unit weight error (UWE) dependence on the magnitude and color of the source means that for consistent estimation, it needs to be re-normalized depending on color and magnitude. RUWE is less certain for objects without color information. \cite{Lindegren_2018} recommends using  RUWE $\leq 1.4$ as a `good' solution. However, since our list is entirely made of bright objects, they will inherently have more noise, so we choose a cut of RUWE $\leq 2.7$ (see also \citealt{Messineo_2019}).

The Bayesian distance estimator {\tt R\_med\_geo} is the recommended package for Gaia stellar distances \citep{Luri_2018,bailer_jones_2021}. Originally proposed in \cite{Bailer-jONES_2015} and further studied in \cite{Astraatmadja_2016},  estimating distances with Gaia becomes an inference problem that is best solved using a simple exponential decreasing space prior. With this method, non-positive parallaxes can still be used, and a bias correction is not required. We use observed parallaxes when estimating the galactocentric coordinates as other position information, such as proper motions, are considered, as suggested in \cite{gaia_coll_2018a}.

\begin{figure}
 \includegraphics[width=\columnwidth]{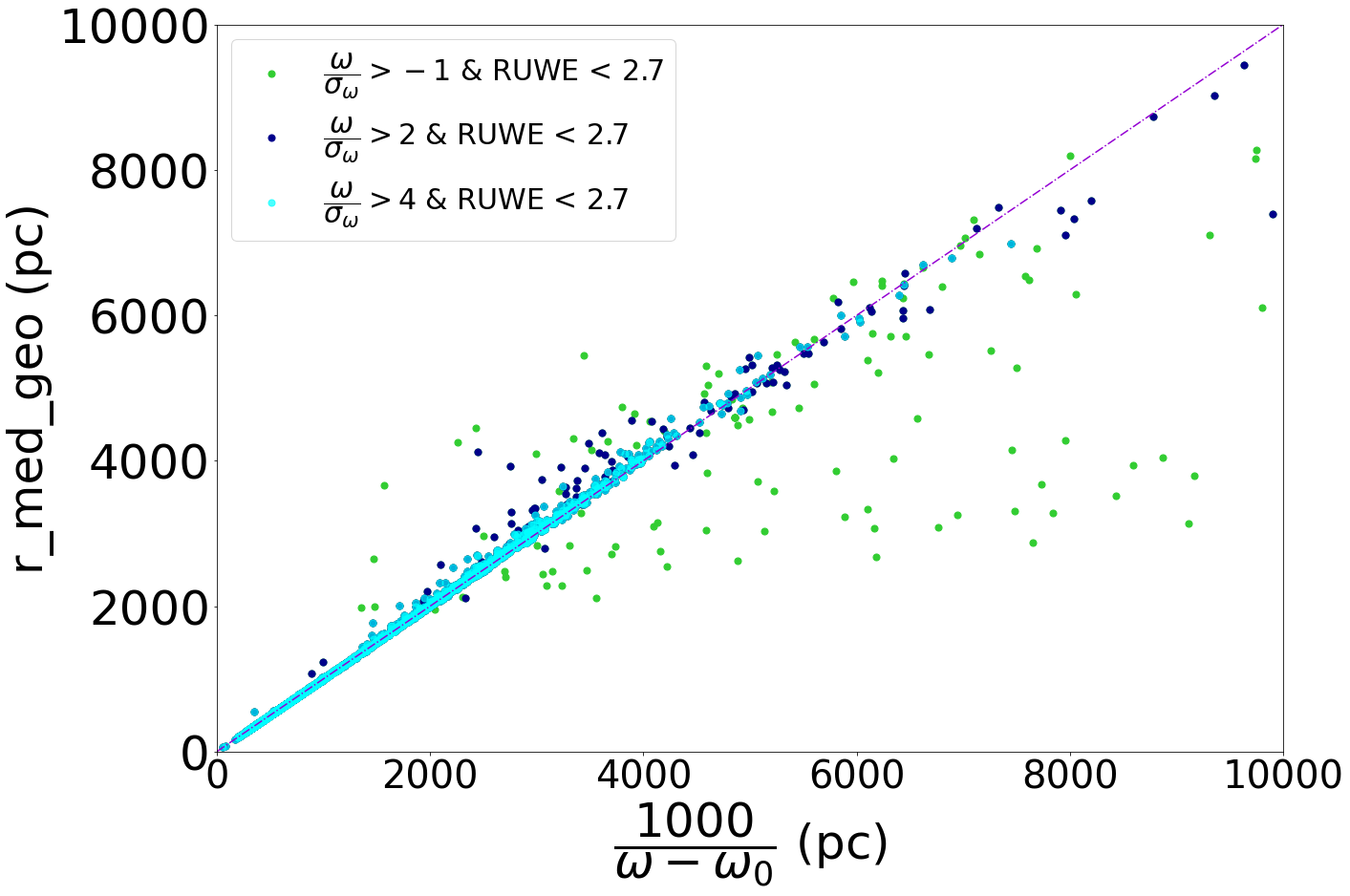}
 \caption{{\tt r\_med\_geo} distance derived by \protect\cite{bailer_jones_2021} using Milky Way model, against parallactic distances from direct inversion of the parallaxes. Those with relative error $>$ -1 and RUWE $<$ 2.7 are in green, and relative error $>$ 2 and RUWE $<$ 2.7 are in dark blue. The cyan objects have relative error $>$ 4 and RUWE $<$ 2.7. The pink dashed line shows the line of zero difference.}
 \label{fig: Distance Cuts}
\end{figure}

Furthermore, distance errors are crucial for us to separate RSGs from impostors. The Gaia collaboration defines two errors: external which includes both random and systematic errors, and internal which is the formal errors reported in Gaia DR3 but only quantify the consistency of measurements\footnote{\url{https://www.cosmos.esa.int/web/gaia/dr2-known-issues\#AstrometryConsiderations}}. The former is the total error, while the latter is not. We calculated the external parallax uncertainties as recommended by applying Eq.~(\ref{external_error}) to the internal uncertainties found in DR3. The external error of the parallax, $\sigma_{\overline{\omega}}$, is defined as
\begin{equation} \label{external_error}
    \sigma_{\overline{\omega}}(ext) = \sqrt{k^2 \times  \sigma_{\overline{\omega}}(int)^2 + \sigma_s^2 },
\end{equation} 
where depending on the G mag, the values of $k$ and $\sigma_s$ are either
$G \lesssim 13 : k = 1.08, \sigma_s = 0.021$ or
$G \gtrsim 13 : k = 1.08, \sigma_s = 0.043$.
With these calculations, we look for relative error, $\rm \overline{\omega} / \sigma_{\overline{\omega}}$, and RUWE such that we retain only objects whose errors will not cause significant uncertainties in later derived quantities. Previous cuts from \cite{Messineo_2019} restricted relative error $>$ 4 to keep the difference between $1/(\omega - \omega_{0})$\footnote{Parallax zero point from \citep{lindegren_2021}} and {\tt r\_med\_geo} to less than $ \rm 5\% $ throughout the entire sample. In an effort to maximize final numbers while balancing quality, comparisons were made between different restrictions, shown in Fig.~\ref{fig: Distance Cuts}. Allowing objects whose relative parallax error $>2$ produces a sample whose maximum difference between distances derived from parallax and {\tt r\_med\_geo} is $>5$\%, but on average, the difference is still $\sim$ 0.7\% with a median value of $\sim$0.2\%, yet almost doubles our sample size in the most favorable regions while preventing the exponentially decreasing space prior to influence {\tt r\_med\_geo} to the degree that it would be dependent on parallax and the Galactic model used by \cite{Bailer_Jones_2018}. 4,863 unique values are retained after applying the distance cut.

\subsection{Photometry}
We utilize near-infrared photometry from the Two Micron All Sky Survey (2MASS). 2MASS finished collecting 25.4 tbytes of data in February 2001, including raw images of 99.98$\% $ of our celestial sphere. The bands measured were the near-infrared  (NIR) J (1.25 $\mu$m), H (1.65  $\mu$m), and K\textsubscript{s} (2.16  $\mu$m) of which errors were calculated and produced for the data set which includes images from almost the entire sky. All images were taken through either Mount Hopkins, Arizona's and Cerro Tololo, Chile's 1.3m diameter telescopes both with 7.8s time accumulated for each sector \citep{Skrutskie_2006}. 

Using ESA pre-crossmatched best neighbour and good neighbourhood list, available 2MASS Point Source Catalogue \citep{Skrutskie_2006} matches were found for $\rm > 99 \%$ of stars that had a Gaia DR3 ID, providing a near-complete sample of JHK\textsubscript{s} measurements and their associated uncertainties. The K\textsubscript{s} magnitudes range from -2 to 16.3 mag with a median value of 4.3 mag. Where available, we also include 2MASS values for the blue, visual, or red magnitude of the associated optical source.

\subsection{Dust} \label{dust}
In \cite{messineo_2005}, dust extinction of RSGs was calculated by using the extinction power law with an index of 1.9 and the intrinsic colors expected for each spectral type. However, RSGs in general show excess circumstellar extinction to those of OB stars in the same regions \citep{Massey_2005}, causing the extinction ratios to differ from those of the main sequence or other stars. This increase in extinction for brighter massive stars is also seen in \cite{neugent_2020b}, which takes direct measurements of $A_V$ for a sample of RSGs in M31.

While a linear relationship between intrinsic color, $(J-K)_0$, and effective temperature works well for RSGs, metallicity in the models causes discrepancies as well as unlikely large extinction values. We also have to take into account that a large portion of our sample's $T_{\rm eff}$ is based on a mean spectral type. Deriving dust based on $T_{\rm eff}$ and intrinsic magnitude means not only would we need to propagate the uncertainties of $T_{\rm eff}$ and observed photometry, but we would also have to deal with the dependence between the two when calculating the uncertainties of intrinsic magnitudes. This method would keep us from treating our uncertainties as uncorrelated, complicating our error propagation. For these reasons and those described in the next section, we use Galactic dust maps to estimate reddening while keeping the extinction ratios from \cite{messineo_2005}.

\subsubsection{3D Dust Map}

In recent years, the development of interstellar dust extinction maps has improved to 3 dimensions spanning many kiloparsecs. We use the combination {\tt mwdust}\footnote{\url{http://github.com/jobovy/mwdust}} created by \cite{bovy_2015} and later updated to replace \cite{green_2015} with \cite{green_2019}. While there are others based on modeling stellar photometry \citep[e.g.,][]{marshall_2006,sale_2014,green_2019}, which would be ideal for our use with 2MASS photometry, none provide results for the entire sky that we need. 

The combination of \cite{bovy_2015} starts with \cite{marshall_2006} based on 2MASS passbands and augments with \cite{drimmel_2003} based on dust distribution fit to COBE DIRBE data. \cite{bovy_2015} unified the projection of \cite{marshall_2006}, \cite{green_2015}, and \cite{drimmel_2003} to HEALPix; note however the resolutions are variable. Collectively, \cite{bovy_2015} provides full sky coverage and preserves small-scale structures in the dust extinction.

As our sample focuses on RSGs, we modified some constants used to transform between bands which are used in the maps so that they fit the extinction transformation equation found in \cite{messineo_2005}, specifically
\begin{equation}
    \frac{A_{K_s}}{A_V} = 0.092, \ \frac{A_H}{A_V} = 0.153, \ \frac{A_J}{A_V} = 0.263
\end{equation}
\begin{equation}
    R_V = \frac{A_V}{E(B-V)}
\end{equation}
where $\rm R_V = 3.1$. Massive stars, however, can have additional attenuation due to local extinction. For example, studies have found values of 3.6 \citep{lee_1970,McCall_2004} or 4.2 \citep{Massey_2005}, and \citeauthor{Levesque_2005_17_ref_15}'s \citeyearpar{Levesque_2005_17_ref_15} survey of Galactic RSGs suggests 4.4. In the absence of excess circumstellar medium (CSM) or other peculiar reddening law, 3.1 still gives good agreement with near-UV for Galactic RSGs \citep{Massey_2005}. Additional dust is a major source of systematic uncertainty, but we can obtain some estimates of their impacts. In the sample of Galactic RSGs from \citealt{Levesque_2005_17_ref_15}, approximately 15\% of those with estimations for V band extinction from spectrophotometry differ significantly---here quantified as impacting our classification of a star as a RSG or not (see \S\ref{contamination section})---from estimations based on $\rm R_V = 3.1$. While their use of moderate-resolution SEDs significantly reduces the chance of discounting high-extinction RSGs, dust attenuation in the NIR is less than that of the optical or lower wavelengths, minimizing our sensitivity to extinction estimations and improving our ability to retain RSG with excess reddening. For these reasons, we conservatively use $\rm R_V = 3.1$, giving us a final sample of stars with either approximately correct or underestimated luminosity. While this makes our sample biased against RSGs with extra extinction, any higher value of $\rm R_V$ would only increase the luminosity of stars and hence increase our sample size.

As \cite{marshall_2006} uses $\rm A_{K_s} = 0.089 A_V$, we make the appropriate correction to match the above. The \cite{green_2019} map is intended to provide reddening in a similar unit as $\rm E(B-V)_{SFD}$ \citep{schlegel_1998} which can be covered to $\rm E(B-V)$ by multiplying 0.884, the constant found in \cite{schlegel_1998} and then moving through the equations above to get $\rm A_{Ks}$. While \cite{drimmel_2003} is normalized to $\rm E(B-V)_{SFD}$, {\tt mwdust} appropriately converts to $\rm A_V$.
 
Since {\tt mwdust} does not provide uncertainties, we estimate them by comparing different maps. We used {\tt dustmaps}, a python package that can be used to query several commonly used maps of interstellar dust, including 2D maps such as \cite{schlegel_1998}, \cite{plank_2014}, \cite{lenz_2017}, and 3D maps such as \cite{marshall_2006} and \cite{green_2015}. If \cite{marshall_2006} was the map with the largest impact on final {\tt mwdust} extinction value, we pulled the uncertainties from {\tt dustmaps} as they were included in the {\tt dustmaps} module. For \cite{green_2019}, values were not directly provided by {\tt dustmaps}, so we followed the method in \cite{green_2015} and took the uncertainty to be half the difference between the 84th and 16th percentiles. While \cite{drimmel_2003} does not provide uncertainties, they point to \cite{lopez-corrediora_2002} who estimates the uncertainties in the K\textsubscript{s} band extinction to be less than 0.015, so all objects where \cite{drimmel_2003} largely determines extinction value have the error set to 0.015 for simplicity. Even though this reduces the confidence of calculated uncertainty for relevant objects, it is $\rm <25\%$ of the final sample. These uncertainties also required the matching conversion applied to the extinction above. 

The values derived from 3D Galactic dust extinction maps are noticeably different from those calculated using intrinsic colors from $\rm T_{eff}$. We summarize the details in \S\ref{summaryofdustmethods}.

\section{Results} \label{3:results}
We determine our most probable RSGs by using the parameter space of $\rm \log(L/L{\textsubscript{\(\odot\)}})$ vs $\rm T_{eff}$. To this end, we first determine the luminosity and $\rm T_{eff}$, then compare them to simulated evolutionary tracks for massive stars, and finally determine the most likely regions by considering the contamination from AGB stars. 

\subsection{Luminosity}
Using the extinction method described in \S\ref{dust}, we estimate the absolute magnitude in the K\textsubscript{s} band based on the de-reddened 2MASS K\textsubscript{s} magnitude and the distance moduli derived from the \cite{Bailer_Jones_2018} geometric distance estimates. Bolometric corrections determined by \cite{Levesque_2005_17_ref_15} were matched to stars based on their spectral type and then applied to the absolute K\textsubscript{s} magnitude, yielding the bolometric magnitude. From the bolometric magnitude, we determined bolometric luminosities relative to the sun, $\rm \log(L/L{\textsubscript{\(\odot\)}})$.  By design, all uncertainties can be treated as independent, so we propagate errors of bolometric luminosities using those of $\rm T_{eff}$, 2MASS photometry, dust extinction, and geometric distance estimates.

\subsection{Comparison to Evolutionary Tracks} \label{evol tracks}

\begin{figure}
	\includegraphics[width=\columnwidth]{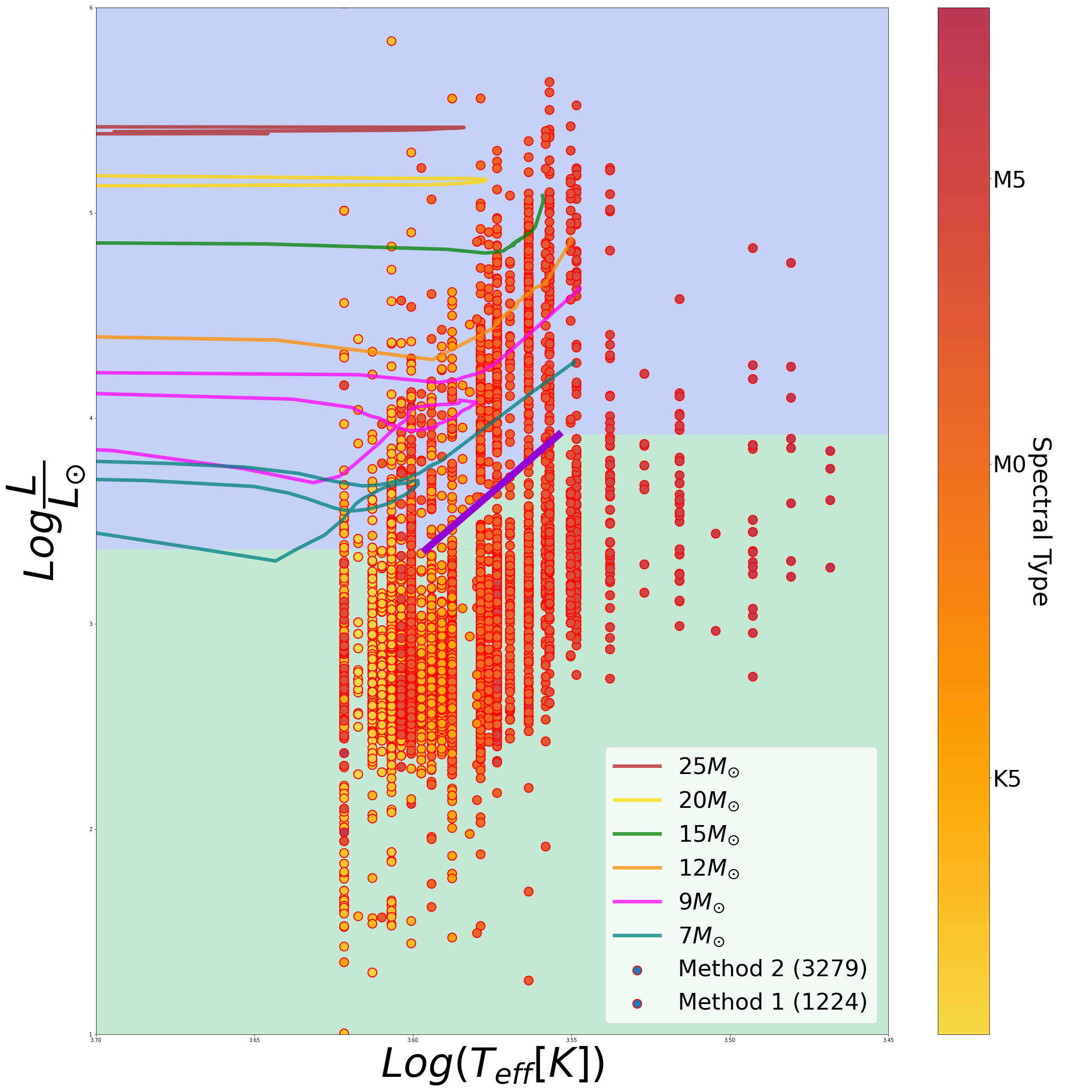}
    \caption{Luminosity vs. ${\rm T_{eff}}$ of stars in both methods, colored by their intrinsic $\rm K_s$ magnitude. Stellar tracks from models at solar metallicity and including rotation from \protect\cite{Ekstrom_2012} are shown for different ZAMS masses, as labeled. The thick purple diagonal line recreates the ascending track from the 7 M\textsubscript{\(\odot\)} progenitor, shifted to account for uncertainties in $\rm \lesssim 7 - 9 \, M_{\odot}$ mass stellar tracks. \label{fig:Stellar Tracks}}
\end{figure}

With $T_{\rm eff}$ and luminosity, we can compare the distribution of stars in the H-R diagram against evolutionary tracks predicted for massive stars. In Fig.~\ref{fig:Stellar Tracks}, we show our sample against tracks of \cite{Ekstrom_2012} for rotating stars computed at solar metallicity ($\rm z = 0.014$) over the mass range  7--25 $\rm M\textsubscript{\(\odot\)}$. For solar metallicity, anything above 25 $\rm M\textsubscript{\(\odot\)}$ is not cool enough to truly be called RSGs, and anything below 7 $\rm M\textsubscript{\(\odot\)}$ would be too small. While these evolutionary tracks do not account for binary evolution, the tracks are still appropriate as binary evolution does not greatly affect the location of RSG on the HR-diagram; \cite{leve_book} succinctly showed this in their Fig.~8.3 which shows a comparison between the single-star Geneva and BPASS binary evolutionary tracks \citep{eldridge_2016,eldridge_2017} (Fig.~7 of \cite{levesque_2018} also supports this). 

The tracks for stars of 7 and 9 $\rm M\textsubscript{\(\odot\)}$ are particularly uncertain as their evolutions are not well understood, and this can be seen as these lines cover wider ranges in luminosity and are different shapes from those of higher mass. If these lines are used as cuts, it is possible that we would exclude true RSGs from our list. To take this uncertainty into account, we follow \cite{Messineo_2019} and use a diagonal cut base on the 7 $\rm M\textsubscript{\(\odot\)}$ track but shifted down and right to the ascending track. This line, plotted in thick purple in Fig.~\ref{fig:Stellar Tracks}, gives us a limit for how faint or red a star could be and still be a candidate RSG: 
\begin{equation} \label{purpleline}
    \log(L/L_\textsubscript{\(\odot\)}) = 51.3 - 13.33\times \log(T_{\rm eff})
\end{equation}
where $\rm \log(T_{\rm eff})$ varies from 3.54 to 3.6 (or M4 to K1, based on \citealt{Levesque_2005_17_ref_15}). As RSGs candidates should be on or between the stellar tracks, with uncertainties taken into account, we can define approximate regions for RSGs. In the following, we determine this in more detail by also considering contamination.

\subsection{Identifying regions} \label{contamination section}

    \begin{figure*}
        \centering
        \begin{multicols}{2}
   {\includegraphics[width=\linewidth]{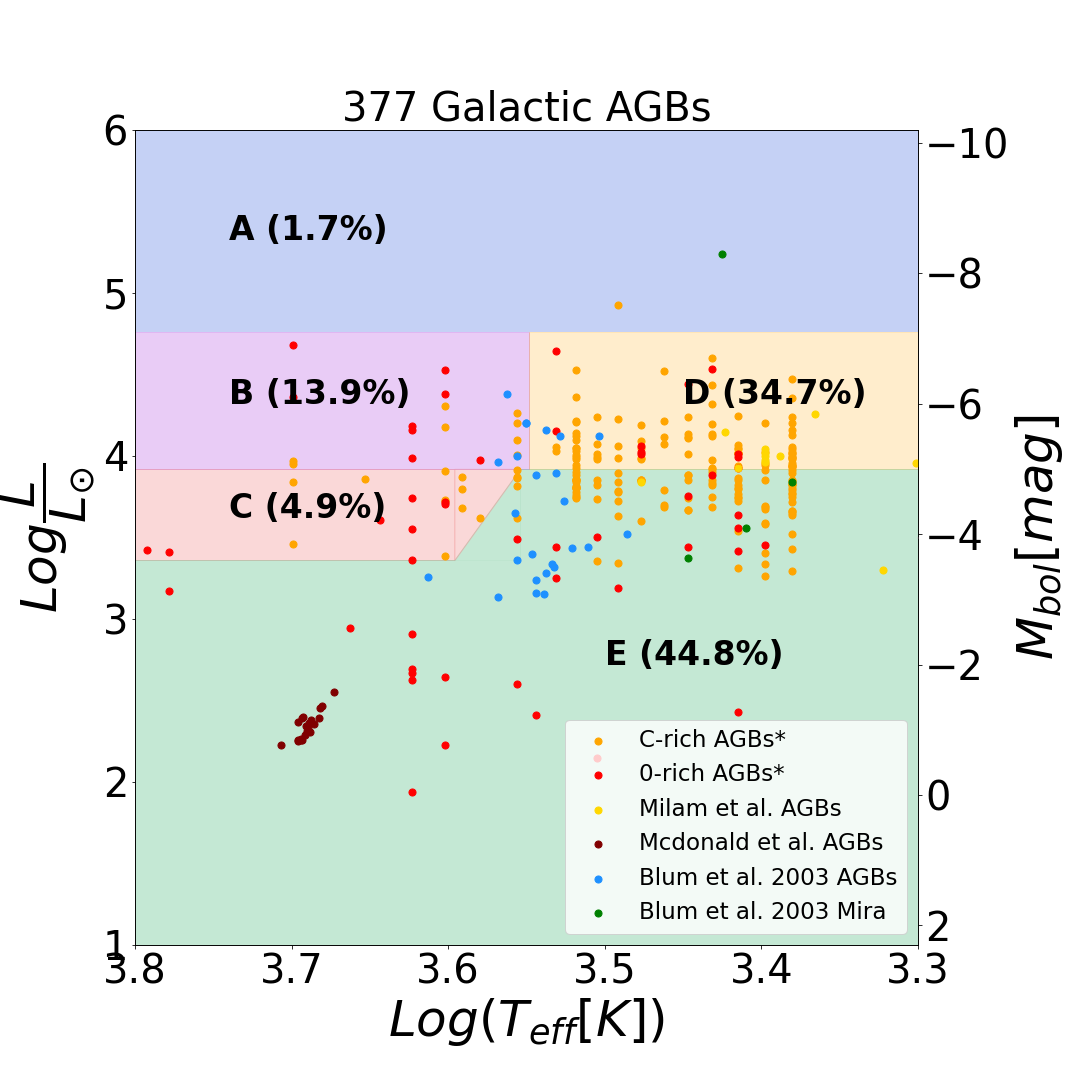}}\par 
   {\includegraphics[width=\linewidth]{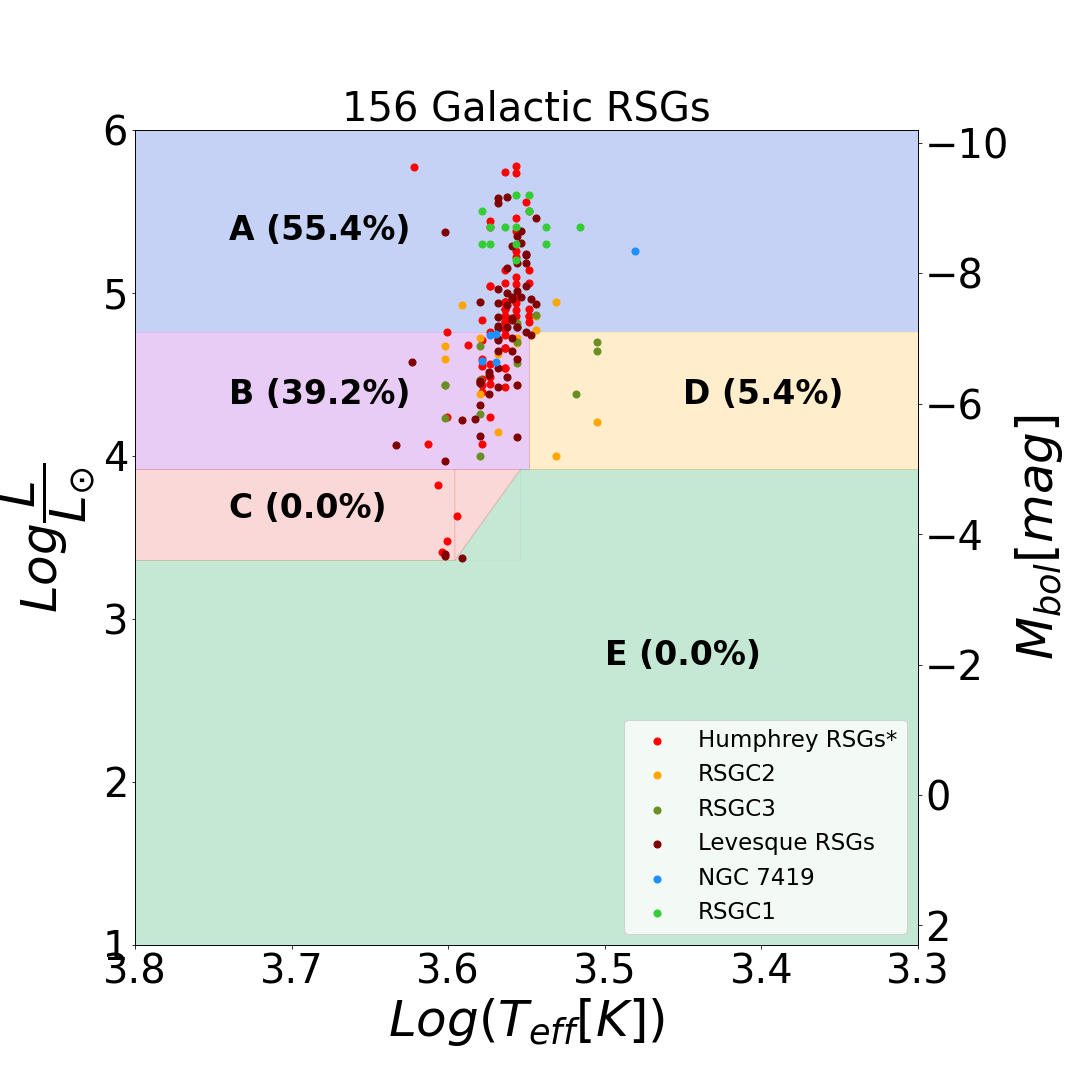}}\par 
        \end{multicols}
    \caption{Left: 377 Galactic AGB, including a set of extreme AGBs from the literature \citep{Mcdonald_2011,Blum_2003_ref_1,goenewgen_2022,milam_2009}; see \S\ref{contamination section} for further details. In each of the regions A--E, we label the percentage of the AGB sample residing in that region. Right: our 156 reference RSGs \citep{figer_2006_ref_46,negueruela_2012_ref_9,Clark_2009_ref_3,marco_2013_ref_39,Levesque_2005_17_ref_15,humphreys_1978_ref_13}, compiled under the guidelines set in \S\ref{gaiabased}.  
    The percentages indicate the fraction of the reference RSG residing in each of the regions A--E. 
    \label{fig:AGB VS RSG}}
    \end{figure*}

The main contamination for RSGs is AGB stars. AGBs have lower initial masses and spend more time on the main sequence; however, in their extreme cases, those that are oxygen-rich or carbon-rich increase in luminosity at a warmer temperature than the average AGB and will cross into the areas that we deem regions for RSG. To quantify this, we compare the distribution of Galactic AGBs and RSGs in the HR diagram. We sourced from the literature Galactic AGBs surveys that include values for both $\rm T_{eff}$ and Luminosity (or $\rm M_{bol}$) \citep{Mcdonald_2011,Blum_2003_ref_1,goenewgen_2022,milam_2009} to quantify how much each region is contaminated. We include a list of extreme AGBs \citep{goenewgen_2022}, including both C-rich and O-rich, as these would most likely be incorrectly selected as RSGs. The Extreme AGBs list includes a small fraction of AGBs supplemented from the Magellanic Cloud. Altogether, we compile 377 AGBs to compare to the reference RSG sample compiled in \S\ref{gaiabased}.

\begin{table}
\centering
\caption{Percentages of the reference RSGs and AGBs in the five regions labeled A$\sim$E. For each region, the percent of AGBs known to be extreme, either carbon or oxygen, is also stated in parentheses. Note that percentages in region C are likely severely underestimated as there is a bias against confirming RSGs or AGBs (see \S\ref{contamination section}). }\label{percentages table}
\begin{tabular}{l|c|c} 
    \hline
    Region&AGBs (Known Extreme AGBs)& RSGs\\
    \hline
    A (Blue)  & 1.7\% (80\%)&55.4\% \\
    B (Purple) & 13.9\% (87.5\%) & 39.2\%\\
    C (Red)  & 4.9\% (89.0\%)& $\sim$0\% \\
    D (Orange)& 34.7\% (0.0\%) & 5.4\% \\
    E (Green) & 44.8\% (38.8\%)& $\sim$0\% \\
    \hline
\end{tabular}
\end{table}

The results are shown in Fig.~\ref{fig:AGB VS RSG}. Based on these, we define five regions. The percentages of AGBs and RSGs in these regions are given in Table \ref{percentages table}. The five regions are:

\begin{figure*}
 \includegraphics[width=1.5\columnwidth]{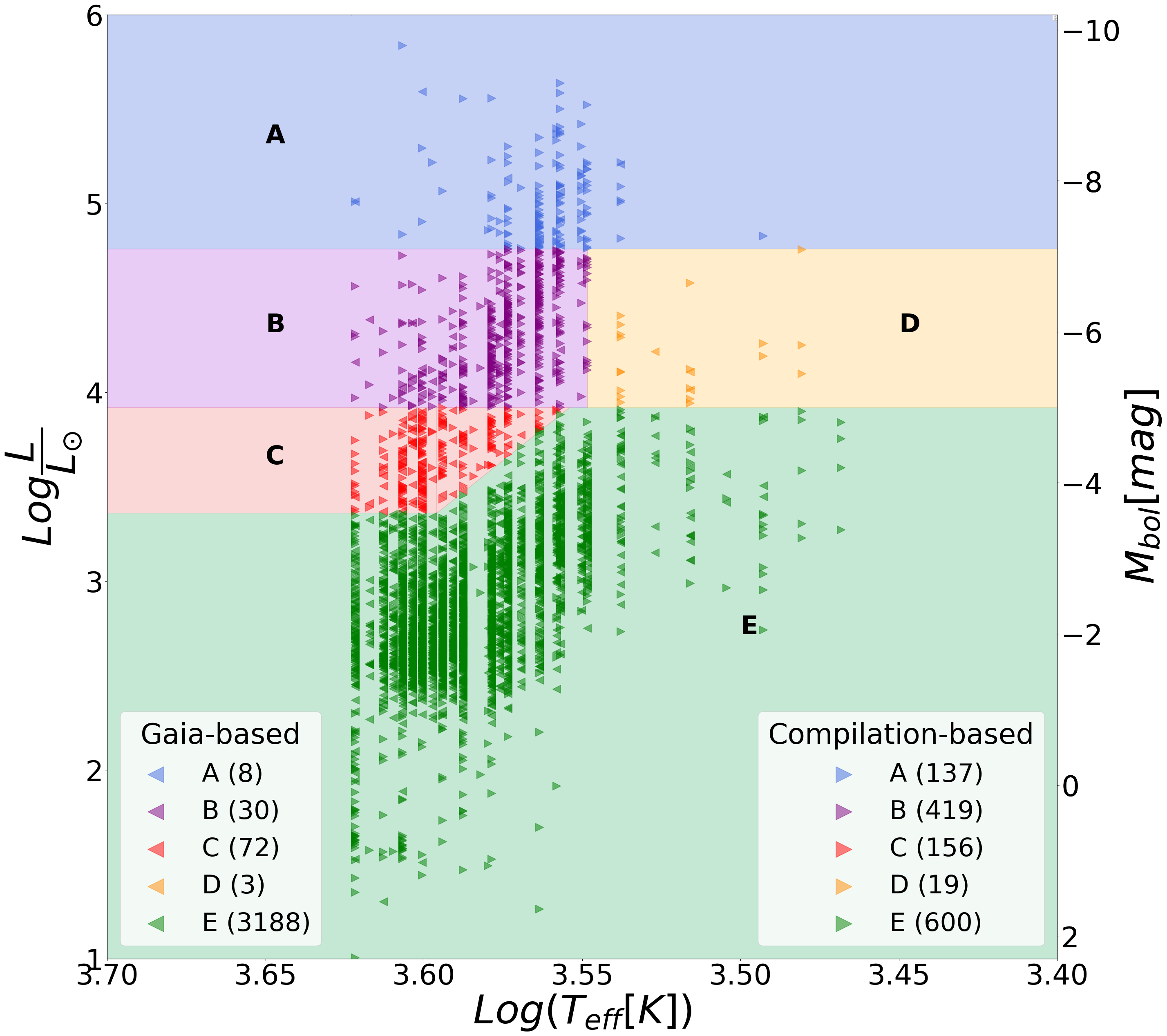}
 \caption{Luminosities vs.~$ \rm T_{eff}$ values of the compilation-based method (arrows pointing right) and the Gaia-based method (arrows pointing left), for stars with relative error $>$ 4 and RUWE $<$ 2.7. Stars within Region A (blue) and Region B (purple) are highly probable RSGs. Region B contains some contamination from AGBs. Region C (Red) contains intermediate-mass AGBs and late M and K type stars with solar masses around 9$\rm M_{\odot}$. As discussed in \S\ref{contamination section}, lower mass RSGs should be in this region. Region D (Orange) and Region E (Green) stars are either too red or too dim to generally be RSGs of any mass, so they are unlikely regions to find RSGs. The number of stars for each region, not accounting for repeats between methods, is shown in the legend in parentheses.}
 \label{fig:areas}
\end{figure*}

\begin{itemize}
\item \textbf{Region A (Blue)}: defined by $\rm log(\frac{L}{L_{\odot}}) \geq 4.76$. It has little contamination: only 1.7\% of the reference AGBs are in region A, while 55.4\% of the reference RSG population is here. Compared to stellar tracks, region A contains high mass RSGs with $\rm \geq 15 M\textsubscript{\(\odot\)}$. 
\item \textbf{Region B (Purple)}: defined by $\rm log(\frac{L}{L_{\odot}})$ between 3.92 and 4.76 and whose spectral type is earlier than M4, $\rm T_{eff}  \geq 3535$ K. It is still RSG-rich (39.2\% of the reference RSGs) but has more contamination than region A (13.9\% of reference AGBs). Region B contains intermediate mass RSGs with masses between 15 and 9 $\rm M\textsubscript{\(\odot\)}$, which fall along well-defined stellar tracks.
\item \textbf{Region C (Red)}: defined by $\rm log(\frac{L}{L_{\odot}})$ between 3.36 and 3.92 and is bluer than Eq.~(\ref{purpleline}). It contains intermediate-mass AGBs (4.9\% of AGBs), and as discussed below, lower mass 9$\rm M_{\odot}$ RSGs should be in this region but biased against observation (0\% of RSGs).  
\item \textbf{Region D (Orange)}: defined by $\rm log(\frac{L}{L_{\odot}})$ between 3.92 and 4.76 and anything $\rm T_{eff} \le 3535$ K. Compared to region B it contains a larger fraction of AGB contamination (34.7\% of reference AGBs) and a smaller fraction of RSGs (5.4\% of reference RSGs). \item \textbf{Region E (Green)}: defined by $\rm log(\frac{L}{L_{\odot}})$ between 3.36 and 3.92 and redder than Eq.~(\ref{purpleline}) or anything with $\rm log(\frac{L}{L_{\odot}}) < 3.36$. It contains a large fraction of the reference AGBs (44.8\%) and none of the reference RSGs. Region E stars are either too red or too dim to be RSGs of any mass. 
\end{itemize}

The reference RSGs along with our reference AGBs illustrate an issue in massive star studies, which is that we lack stellar tracks certain enough and photometric characteristics that could distinguish between 7-9$\rm M\textsubscript{\(\odot\)}$ RSGs and contamination from lower mass stars and extreme AGBs \citep{siess_2006,Poelarends_2008}. The best options beyond modeling and photometry also fail in certain cases; observation of spectral lines sensitive to surface gravity fail for cool emission spectra, high-resolution follow-up is not realistic logistically for large data sets, and use of maser emissions requires circumstellar masers to distinguish between supergiant and giant stars \citep{leve_book}. Even in this list of well-studied RSGs, almost none have masses below 9 $\rm M\textsubscript{\(\odot\)}$ even though these should be more common than higher-mass RSGs. This results from the prevailing selection criteria, similar to that in \cite{massey_olsen_2003}, which applies a conservative lower limit and filters out lower mass RSGs. Ultimately, we similarly remove Region C even though it should contain low-mass RSGs.

\subsection{The Catalog}

We show in Fig.~\ref{fig:areas} our five regions and our sample combining the compilation-based and Gaia-based methods. We keep stars in Regions A and B for our final RSG list. This results in 134 and 436 stars, respectively. However, we can also include stars that satisfy the region cuts after consideration of errors. We include distance, photometry, dust extinction, and effective temperature as errors. These stars are listed in their original region identifier, meaning C, D, or E (see Appendix \ref{appendixList} for a complete column listing of the catalog). Still, they are included as ``likely'' RSGs in the final catalog due to uncertainties. This adds another 62 stars, primarily from Region C. 

\subsection{Characteristics Flags} \label{chara section}
We add additional flags that give extra insight into how likely an object is an RSG, going beyond the required luminosity and temperature cuts. The statistics of these flags are shown in Table \ref{table: characteristics of stars breakdown}.

\subsubsection{Clusters}
Due to the dense molecular cloud that massive stars are born in, RSGs often form as part of a cluster or OB associations, so RSGs searches are often restricted to a specific cluster. We thus search the literature to see if each star is a member or candidate member of any Milky Way cluster. We include the cluster name, any radial velocity derived from cluster dynamics, and any cluster details, like age, approximate distance, and mass, into the catalog if available. The main clusters found include Stephenson 2 (RSGC2), PER OB1, and CAR OB1. 

\subsubsection{Multi Star Systems}
Like OB associations and clusters, the dynamics of the molecular clouds are appropriate for forming binary systems. Considering the significant binary fraction for main-sequence massive stars, once a massive star reaches mass transfer at the Hayashi limit, it is likely that this will cause an increase in the binary fractions of massive stars. To what extent it increases depends on the theoretical treatment of RSGs. It is almost certain that the limitation of observations and the rarity of RSGs are the causes of so few known binaries. 

We thus include a flag for stars that have evidence of interactions with a companion. The catalog contains a column to flag binary and multi-star systems and a column detailing if the spectra show a binary (Spectral Binary), if Gaia DR3 has identified it as a multiple object system ({\tt non\_single\_star}), or if the literature defines it as such. There is also a corresponding reference column that will give the relevant reference.	

\subsubsection{Variability}
RSGs are expected to be radial pulsators with luminosity-dependent pulsational periods \citep{Stothers_1969,stothers_1972,Heger_1997,Guo_2002} leading to photometric variation. Mass-loss rates have been measured up to $\rm 10^{-5}$ $\rm M\textsubscript{\(\odot\)}$$yr^{-1}$ \citep{Beasor_2016_7} and are another driver for variability. 
Spectroscopic variability is affected by the dynamics of their outer layers, like extended atmosphere and optical depth effects near the Hayashi limit or convection, pulsation, and mass loss. While the dominant driver responsible for this variation is unknown, it is an important characteristic of RSGs. We, therefore, included a flag for any object which is designated as a variable star, either in the literature, AAVSO, ASAS-SN \citep{shappee_2014,jayasinghe_2020}, or through Gaia DR2/DR3's photometric variability flag. The latter uses statistics, time-series, and data mining analysis on Gaia $\rm G$, $\rm G_{BP}$, $\rm G_{RP}$ photometry as well as parallaxes and positions
\citep{Holl_2018_gaia_dr2_variability} and using statistical and machine learning methods built from a global revision of major published variable star catalogs \citep{Eyer_2022_gaia_dr3_variability}).

\begin{table*}
 \caption{Breakdown of characteristics for highly probable RSGs included in the final catalog separated into Region A (138 stars) and Region B (460 stars). Close RSGs are those whose proximity, extended radius, and/or brightness lead to poor or no measurements in Gaia and were manually added to our catalog as described in \S \ref{not gaia}.}
 \label{table: characteristics of stars breakdown}
 \begin{tabular}{lcccccccc}
    \hline
    Region&OB Association&Variability&Binary& Measured Magnetic&Runaway\\
    &\& Clusters&(Spectra or Photometry)&(Spectra or Photometry)& Field&\\
    \hline
    Region A & 37& 137& 11& 1&1\\
    Region B & 62& 312 & 60&2 &0\\
    Close RSGs & 7& 12& 4 & 10&3\\
    \hline
 \end{tabular}
\end{table*}

\section{Discussion  and interpretations} \label{discussion and interperation}

\subsection{Specific objects}

\subsubsection{RSGs which are not in Gaia DR3} \label{not gaia} 

Even though Gaia has a large magnitude range G $\sim$ 3 to 21 mag, for any star with G mag $< 5$, the brightness could lead to saturation and subsequent systematic errors in parallax. This happens since Gaia astrometric measurements use the stars' magnitudes in part to self-calibrate, so this becomes a problem as bright stars are scarce \citep{Lindegren_2018}. Also, the extreme widths of massive stars can limit the accuracy of Gaia's parallax measurement. For stars with magnitudes bright and radii wide enough to have large enough errors to be removed from our pipeline but are generally accepted as RSGs in literature, we reintroduce them into our sample. There are 12 stars in this category. It includes $\theta$ Del, $\mu$ Cep, S Per, VX Sgr, and NML Cyg who are in Gaia DR3 but whose measurements are saturated and accuracy is reduced below our threshold, and alf Ori (aka Betelgeuse), alf Her, alf Sco, eps Peg, $\zeta$ Cep, and lam Vel who are not in Gaia DR3 due to some combination of their distance, brightness, and radii. We also include VY CMa, whose brightness and distance should not lead to over-saturation but is excluded as its radius, one of the widest RSGs in our galaxy, hinders accurate distance measurements. 

As the above list is not measured with Gaia DR3, we do not use our pipeline to determine their bolometric magnitude or luminosity. Instead, we used values determined throughout the literature. These papers are denoted in the catalog, and a full list can be found in Appendix \ref{list of close star ref}). While we tried to fill as many columns as possible for each non-Gaia object, this was not possible for some columns, especially those from Gaia. A list is provided in a single reference column for objects whose values come from multiple sources.

\subsubsection{MY Cep}
The use of stellar tracks and reference RSGs/AGBs to determine our regions for possible candidates resulted in MY Cep being removed from our list, not because it does not have a high enough luminosity, but because its $\rm T_{eff}$ is too cool. However, MY Cep is generally acknowledged as a RSG \citep{Levesque_2005_17_ref_15,MAURON_2011,Humphreys_2020} and it makes sense to keep it when considering the uncertainty in $\rm T_{eff}$.

\subsection{Runaway RSGs}
\cite{blaauw_1956a,blaauw_1956b} discovered a sample of OB stars with significantly higher space velocities than their surroundings, denoted runaway stars (RWs). RWs were first theorized to originate from binary systems whose primary undergoes SN and shoots the secondary out, creating their peculiar 
velocities. However, this has been disproven as a primary mechanism \citep{gies_1986}. Other explanations hypothesize their existence from dynamical ejection \citep{leonard_1990} or interactions with massive black holes \citep{capuzzo_2015,fragione_2016}. Estimates for OB stars with peculiar radial velocities $\gtrsim$ 40 $\rm km s^{-1}$ reach as high as 50\% of the population \citep{gies_1986}. However, few RW massive stars are known. Galactic RW RSGs include those identified due to the presence of bow shocks: Betelgeuse (56 $\rm km s^{-1}$; \citealt{noriega_1997,Mackey_2012}, $\mu$ Cephei (22 $\rm km s^{-1}$; \citealt{cox_2012}), and IRC -10414 (70 $\rm km s^{-1}$, \citealt{gvaramadze_10.1093/mnras/stt1943}) and by Kinematical data from the Gaia DR2 catalog: HD 137071 (54.1 $\rm km s^{-1}$, \citealt{comeronn_2020,tetzlaff_2011}) which interestingly is both the only known K-type and whose proposed velocity evolution favors the nondominant mechanism of ejection due to SN \citep{comeronn_2020}). All of these are included in the final sample.

\subsection{Presence of Magnetic field}
While the extent to which magnetic fields can contribute to total mass loss in RSGs is unknown, perturbation of fields with strength greater than several Gauss could have effects through the production of Alfven waves \citep{hartmann_1984A,charbonneau_1995,vidotto_2006}. There have been several Galactic RSGs with measured magnetic fields, including $\sim1$ G in Betelgeuse \citep{auriere_2010,tessore_2017}. Observations of surrounding circular polarization from Zeeman-splitting suggest a surface magnetic field as strong as 4 G for VX Sgr \citep{vlemmings_2005}, $\sim1$ G for 32 Cyg, and $\sim2$ G for $\rm \lambda$ Vel \citep{grunhunt_2010}. \cite{tessore_2017} detected $\sim1$ G from CE Tau and $ \rm \mu$ Cep, and several G in $\rm \alpha^1$ Her similar to the magnetic fields of AGBs. Measurements of VY Cma show disagreement with estimations from $\rm H_2O$ maser observation ranging from 90 to 180 mG \citep{vlemmings_2002} and estimations based on linear polarization suggesting a lower limit of $\sim 10$ G \citep{shinnaga_2017}. Discussion detailing possible reasons for this disagreement are presented in \cite{shinnaga_2017}. Non-detections of post-RSG stars have also been made, such as yellow SG (YSG) $\rm \rho$ Cas \citep{tessore_2017}, which suggests either a very weak magnetic field or its dissipation, giving insight into the stellar evolution of RSGs that evolve back to warmer temperatures. 

\subsection{Spatial Distribution}
\begin{figure*}
        \centering
        \begin{multicols}{2}
   {\includegraphics[width=1\linewidth, height=1\linewidth]{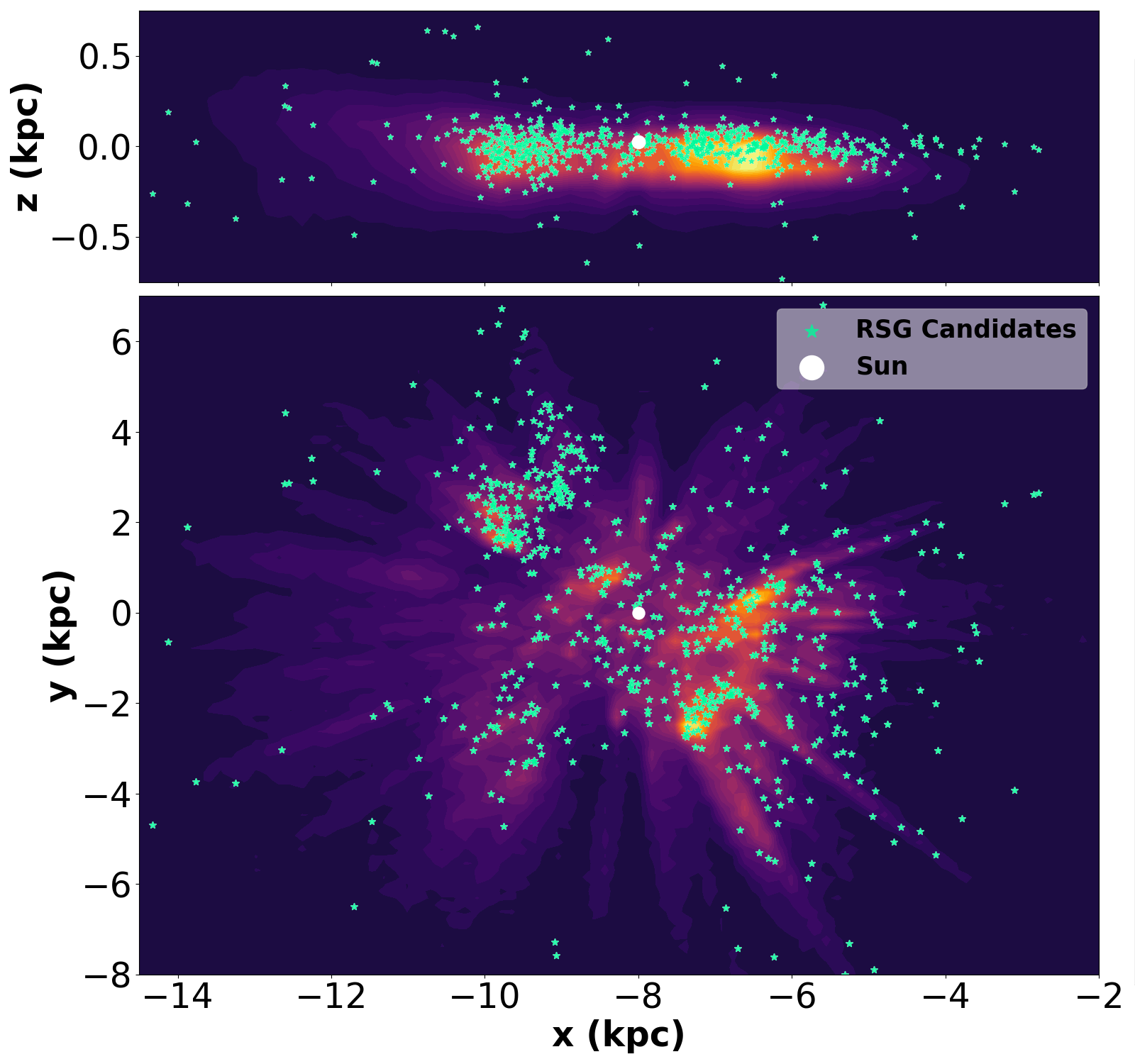}}\par 
   {\includegraphics[width=1\linewidth, height=1\linewidth]{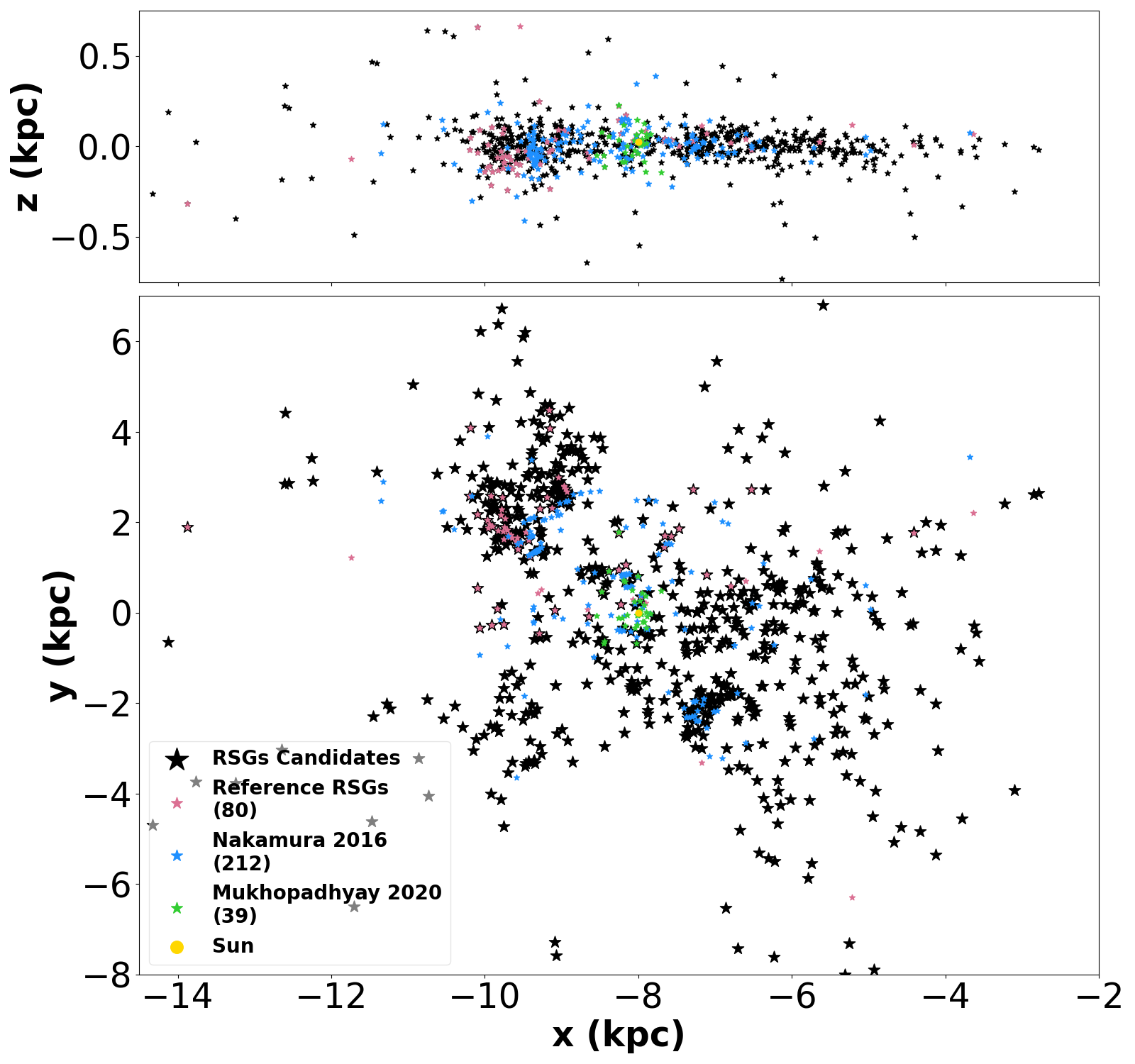}}\par 
        \end{multicols}
    \caption{Left panel: Our final RSGs (teal stars) in Galactocentric coordinates compared to a population of inner Milk Way young stars (shaded based on stellar density from a 2D histogram using cells of 150x150 pc for y vs. x and 150x115 pc for z vs. x) on the upper main sequence selected from Gaia DR3 based on methods of \protect\cite{drimmel_2022} whose density distribution reconstructs the spiral arms of the Milky Way, specifically, Perseus arm, Orion spur, and Sagittarius arm. The Sun's location (8, 0, 0) is marked in white. Right panel: Our RSGs catalog (large black stars) compared to previous studies: \protect\cite{Nakamura_2016} (small light blue stars), \protect\cite{Mukhopadhyay_2020} (small green stars), and reference RSGs (pink). The Sun location (8, 0, 0) is marked in yellow. The number of stars is shown in the legend in parentheses.} 
    \label{fig:spatial_distribution}
    \end{figure*}

The current collection of Galactic RSGs is still highly incomplete, with little known about the total spatial distribution \citep[see, e.g.,][]{Davies_2009_8,Messineo_2016a}. Even with our catalog being the largest collection, it only extends out 12.9 kpc when the radius of the Milky Way is $\sim$ 30 kpc. Figure~\ref{fig:spatial_distribution} shows the spatial distribution of our catalog. In the right panel, we compare with the spatial distribution of other catalogs in the literature: \cite{Nakamura_2016} uses less stringent luminosity cuts but still reaches approximately the same distance, and \cite{Mukhopadhyay_2020} where the inner Galactic Center's RSGs and Blue SGs are represented but only goes out to  $<$ 1 kpc. Note that we see a more prominent dearth of stars around the solar system, driven by the saturation of Gaia photometry (see \S\ref{not gaia}). 

The driving factor for the lack of stars beyond 12.9 kpc comes down to the availability of spectra. Even in \cite{skiff_2014}, one of the largest collections of Galactic stellar spectra, only $\sim0.3$\% meets the most basic spectral characteristics for RSGs. As the determination of spectral type from spectra for RSGs involves either labor, limitations to sample size, or reduction of confidence, RSGs searches have been limited by pointing in the direction of OB associations or clusters, which generally have a higher population of RSGs. However, for inner Galactic supergiants, only $\rm \approx 2\%$ are associated with stellar clusters \citep{Messineo_2017}. The spectra of RSGs are not only observationally limited but also lack well-defined spectral standards. This is especially true of K-type stars, which are often broken into late and early K-type stars rather than subtypes. 

Even after locating candidates, intrinsic characteristics that help determine whether it is a RSG (e.g., pulsation properties and chemical abundances) are not easy to obtain. As discussed in \S\ref{contamination section}, reliable confirmation is also difficult because the colors of RSGs are not unique and match those of giant late-type stars, specifically from low masses to super-AGBs of 9–10 $\rm M_{\odot}$.

These issues all lead to the number of Galactic late-type stars of class I being less than $\sim$1000, and, when not considering our new catalog, $\sim$400 RSGs known throughout various surveys. Major catalogs like \cite{humphreys_1978_ref_13} lists 92, \cite{Elias_1985_ref_14} lists 90, \cite{Levesque_2005_17_ref_15} analyzed the spectra of 62, \cite{jura_1990_ref_12} lists $\sim$135, even though more than $\sim$5000 RSGs are predicted by \cite{Gehrz_1989}.

\subsection{Radii range}
The expanse of RSG radii makes measurements challenging as it limits the observational effectiveness of interferometric determinations and requires sufficiently accurate distances. For the several dozen stars with well-measured radii, comparisons to estimates from $\rm L=4\pi \sigma R^2 T_{eff}^4$ show good agreement \citep{van_belle_2009,wittkowski_2017,wittkowski2012,arroyo_torres_2013,arroyou_torres_2015}. Solving the equation for the radius of Betelgeuse, $\rm \sim855$ $\rm R_\odot$, compares well to interferometric measurements \citep{Dolan_2016} $\rm \sim890$ $\rm R_\odot$, especially as there is variation depending on observed wavelength, treatment of asymmetries, among other things \citep{Townes_2009}. Based on this simple but effective relationship, we provide radii for all stars in our catalog, resulting in a range of $\rm 10^{1.38}$ to $\rm 10^{3.28}$ $\rm R_\odot$.

\begin{figure*}
 \includegraphics[width=2\columnwidth]{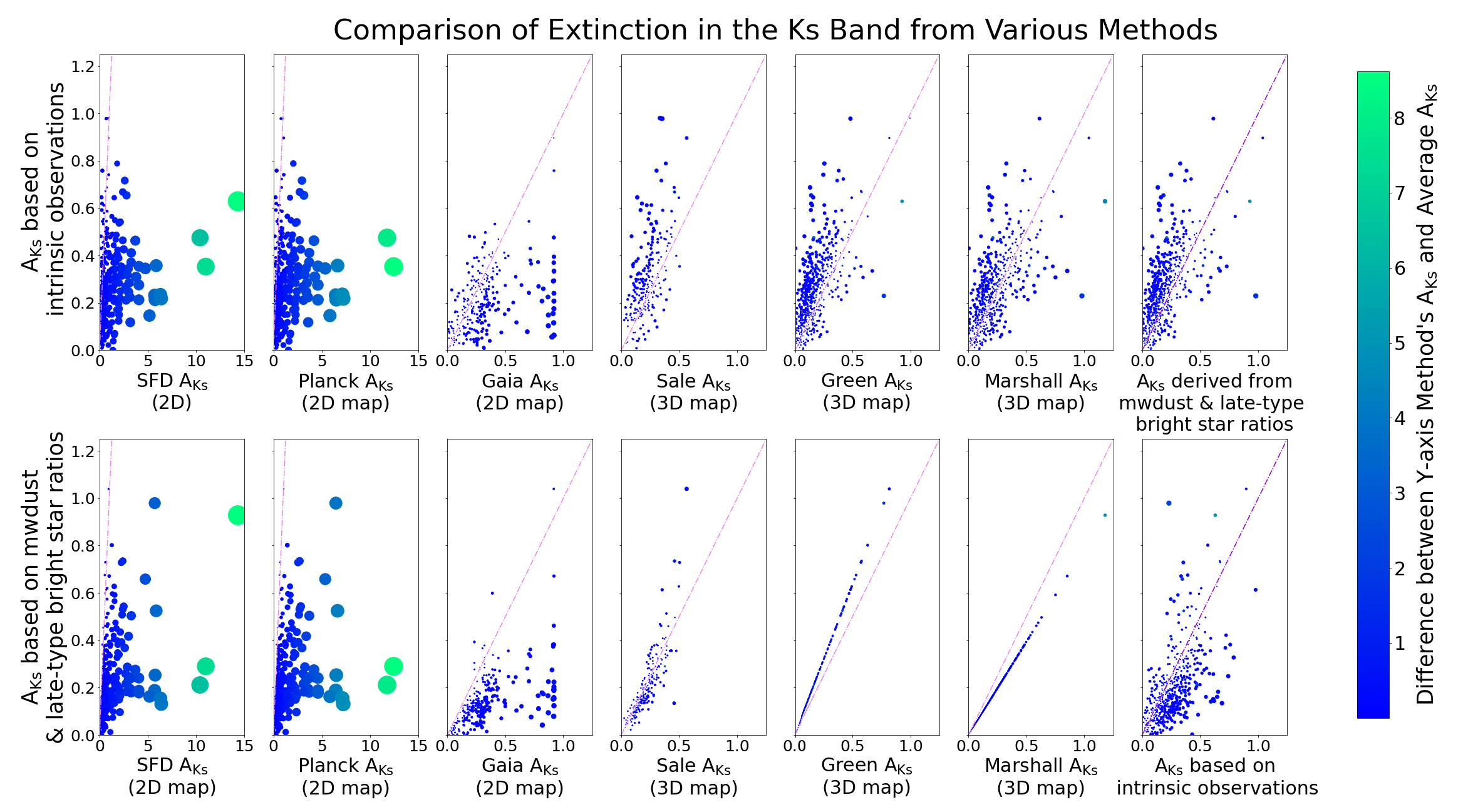}
 \caption{Comparisons of 3D dust maps, {\tt mwdust}, and using ICMS to other methods of determining Galactic extinction. The top row shows ICMS along the y-axis and the bottom row shows {\tt mwdust} extinction converted to the $\rm K_s$ band along the y-axis, with each being compared to 2D maps \citep{schlafly_2011,plank_2014,gaia_dust_2022} and 3D maps \citep{sale_2014,green_2019,marshall_2006} along the x-axes. Points are colored by the size of the difference in the y-axis estimated extinction and the average extinction value across all methods. \label{fig:DUST_distribution_extinction}}
\end{figure*}

\subsection{Mass loss rates}
For massive stars, more than half of their mass loss happens after the main sequence, with RSGs losing between $\rm 10^{-7}$ to $\rm 10^{-3}$ $\rm M_\odot yr^{-1}$ \citep{MAURON_2011,van_loon_2005}. Observations of CSM around type II SNe suggest that mass loss during the final stages of the RSG phase is accelerated. \cite{Forster:2018mib} surveyed 26 SN II within hours of their discovery and obtained optical light curves from High Cadence Transit Survey (HiTS) \citep{martnez-palomera_2018} that when compared to detailed models suggest density profiles consistent with $\rm >10^{-4}$ $\rm M_\odot yr^{-1}$. Confined dense CSM was confirmed to surround SN2013fs that was estimated to be ejected during the final $\sim$1 yr prior to explosion at a rate $\rm \sim 10^{-3}$ $\rm M_\odot yr^{-1}$ \citep{YARON_2017}. However, with the dominant mechanism for significant mass loss unknown and explanations incomplete, derivations of general relations are restricted to observations and limited modeling. While there are some discrepancies between different mass loss rate relations, we estimate the mass loss of our RSG candidates based on \cite{van_loon_2005}, which gives 
\begin{multline} \label{mass loss equation}
    {\rm log_{10} }(\dot{M}) {\rm = -5.5} \\
    \rm + 1.05log_{10} \left(
    \frac{L}{10000L_\odot} \right) - 6.3log_{10} \left( \frac{T_{eff}}{3500K} \right)
\end{multline} 
based on observations of a small sample of M-type stars and the assumption of a dust-driven wind model, most applicable to dust-enshrouded RSGs and oxygen-rich AGB stars. We find a range of $\rm 10^{-7.76}$ to $\rm 10^{-3.87}$ $\rm M_\odot yr^{-1}$ in our sample.

\subsection{Dust Extinction Method Comparison} \label{summaryofdustmethods}

\begin{figure*}
 \includegraphics[width=2\columnwidth]{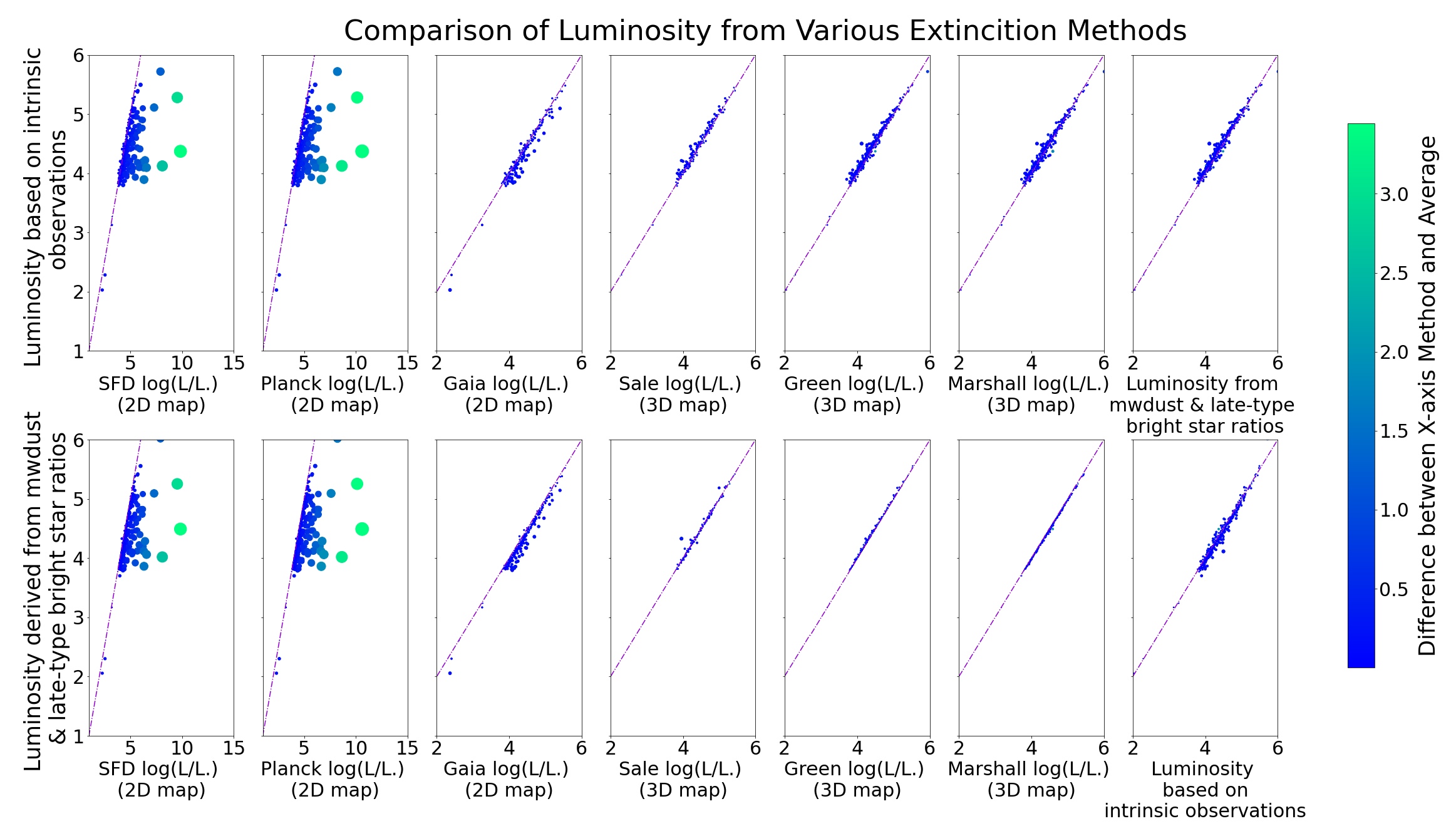}
 \caption{Same as Fig.~\ref{fig:DUST_distribution_extinction} but showing how different dust correction methods impact our determined luminosity.  
    \label{fig:DUST_distribution_lum}}
\end{figure*}

In Section \ref{dust}, we discuss our reasoning for departing from previous methods, which use observationally determined intrinsic colors for different spectral types of massive stars and working backward to determine their dust extinction. However, when a comparison is made with the previous method, our 3D dust map, and other dust maps, we see differences significant enough for us to outline here. We compare six different methods of estimating dust: \cite{schlegel_1998,schlafly_2011,sale_2014,marshall_2006,green_2019,plank_2014,gaia_dust_2022}. We also explore how the calculations of bolometric luminosity would shift depending on the method used for extinction. Figure \ref{fig:DUST_distribution_extinction} shows comparisons of dust corrections, spanning 3D dust maps, {\tt mwdust}, and the use of intrinsic colours of massive stars (ICMS). The close alignment to the 1-to-1 line, shown dashed in pink, for Marshall and Green results from {\tt mwdust} predominately using those two maps for their healpix extinction map. We see that {\tt mwdust} produces a smaller spread across the board and generally has smaller differences from the average values as well. In Fig.~\ref{fig:DUST_distribution_lum} we show how they impact our determined stellar luminosities. Except for the earliest 2D maps of \cite{schlafly_2011} and \cite{plank_2014}, the luminosities determined by our dust method remain very similar to those from more modern dust maps.

We also look into how our final estimates for bolometric luminosity compare to those in the literature; their values are more confident as they generally focus on a small sample of RSG candidates and directly determine their characteristics. Using our reference RSGs from \S \ref{gaiabased}, we compare the changes that would result from using DR2 geometric distance ({\tt r\_est} \citealt{Bailer_Jones_2018}) instead of DR3's. The mean, median, and mode values for the difference between the reference RSG values and those estimated using our method and DR3 are all $\rm \geq15\%$ smaller than those from DR2. The general trend of DR3 based values being closer to their reference values reinforces our use of distances primarily from {\tt r\_geo\_med} but supplemented with {\tt r\_est}. However, both methods systematically underestimate luminosity, with a few outliers generally caused by areas of high dust. Stars which have the required luminosity and $\rm T_{eff}$ are very likely to be bright or brighter than our limits as systematics taken into account would shift them higher, validating the inclusion of stars whose estimated values are not high enough but whose uncertainty places them in one of the two retained regions. 

\section{Comparison with previous works} \label{catalog comparison}

Collecting the Milky Way's RSGs has been a necessary challenge, particularly for two areas of astrophysics: stellar astronomy and studies of massive stars pre-explosion  \citep{Messineo_2019,messineo_2023} and astroparticle focused more on neutrinos and the explosion itself \citep{Nakamura_2016,Mukhopadhyay_2020}. The motivations and approaches behind all efforts vary considerably.

\cite{Messineo_2019} focused on finding candidate RSGs in Gaia DR2. They compiled Galactic spectral catalogs whose objects are in Gaia DR2 to estimate the effective temperature ($\rm T_{eff}$) and determine reliable distance measurements, and match them with infrared and optical measurements from various sources. Their combination of multiple bands with effective temperature allowed them to estimate stellar bolometric magnitude in two ways: by using photometric measurements and by integrating under the SED. This resulted in 889 late-type stars and 43 highly-probable RSGs.

The release of Gaia DR3 and associated data products presented new opportunities to categorize Galactic RSGs. A new list of late-type bright stars, including RSG and bright AGBs was developed using Gaia DR3 Apsis \citep{gaia_dust_2022,gaia_extinction_2013} and Gaia DR3 GSP-Phot and GSP-Spec parameters of known K- and M-type stars of Class I luminosity by \cite{messineo_2023}. In addition to the previous work, there are 203 new entries of late-type bright stars with 15 S-type, 1 S/C, 9 C-rich, and 20 confirmed new RSGs with 6 having bolometric magnitudes brighter than the AGB limit.

For both \cite{Nakamura_2016} and \cite{Mukhopadhyay_2020}, potential progenitors and their properties were compiled from the literature. \citealt{Nakamura_2016} explored the multi-messenger signals of a nearby CCSN, and simulated a complete sequence of signals, detectors' abilities limited their sample to $\sim$3 kpc. For future detection prospects, \cite{Nakamura_2016} compiled 212 Wolf-Rayet stars along with RSG candidates, sometimes with the luminosity class of II or even III. \cite{Mukhopadhyay_2020} heavily focused on the potential of liquid scintillators neutrino detectors to localize presupernova neutrino signals, a task which may be achieved at close distances. Thus, their list is restricted to $\lesssim$ 1 kpc, for a total of 31 containing red and blue massive stars. 
No analysis is done beyond compilation to determine the accuracy of CCSN progenitor status. As the purpose and method for listing possible CCSN progenitors candidates were relatively consistent, there was an overlap of the most nearby candidates.

By comparison, our work resulted in 578 Milky Way candidate RSGs. To obtain this enhancement, we begin by following the method of \cite{Messineo_2019} but modify it in a few key ways to tailor our analysis for use with multi-messenger astronomy. For example, we do not restrict our sample to those in the Galactic plane, and we place an emphasis on having more RSGs at the expense of having slightly more AGB stars contaminating our list, similar to \cite{Nakamura_2016} and \cite{Mukhopadhyay_2020} but to a lesser degree. Furthermore, we work with the updated Gaia DR3, though we do not yet include Gaia DR3 spectra or Apsis parameters as \cite{messineo_2023} does. Our method for determining a single spectral type for objects with multiple designations considers the increase in data and follows a more strict prescription than that of \cite{Messineo_2019}. We also implement a new method that adds 33 unique highly probable and likely RSGs. This method uses a sample of known K- and M-type stars included in Gaia DR3 to determine a Gaia G band photometry cut-off. It is combined with external spectra, working similarly to \cite{messineo_2023}'s use of Gaia Apsis parameters and spectra.

Even though we obtain a larger number of RSG candidates---with potentially more contaminants---it is worth pointing our we are far from complete. Given the Milky Way's core-collapse supernova rate of a few per century \citep{tammann_1994,rozwadowska_2021} and the duration of the RSG phase for the most common $8M_\odot$ star being $\sim 0.5$ Myrs \citep{eldridge_2008}, the number of RSGs at any given time in the Milky Way is approximately $\sim 5\times10^5 / 100 \sim 5000$ \citep{Gehrz_1989}.

\section{Multi-messenger Astronomy} \label{MMA}

Multi-messenger astronomy is based on coordinating observations across multiple messengers---from electromagnetic radiation to gravitational waves and neutrinos---and on the interpretation of the joint results to optimize the science attained. With coordinated efforts, both detections and non-detections lead to useful information on the physics behind different astrophysical phenomena. A key to successful coordination for transient events such as CCSNe is in aligning the observatories in time. To this end, advanced warning of an impending CCSN is highly useful. 

\begin{figure*}
    \centering
    \begin{multicols}{2}
   \subcaptionbox{Calculations based on CD-34 11794 (263$^{\circ}$, -34$^{\circ}$, and 4.1 kpc) as the progenitor. 15 stars lie within the 90\% credible region with 2 of those matching the combined distance estimation and angular resolution.\label{fig:1}}{\includegraphics[width=1\linewidth]{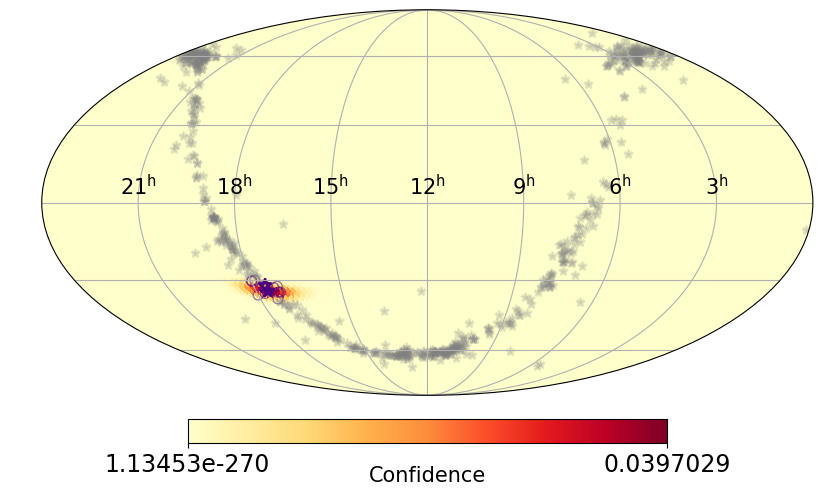}}\par
   
   \subcaptionbox{Calculations based on HD 303250 (161$^{\circ}$, -58$^{\circ}$, and 2.4 kpc) which lies within cluster CAR OB1 as the progenitor. 85 stars lie within the 90\% credible region, with 20 of those matching the combined distance estimation and angular resolution. \label{fig:2}}{\includegraphics[width=1\linewidth]{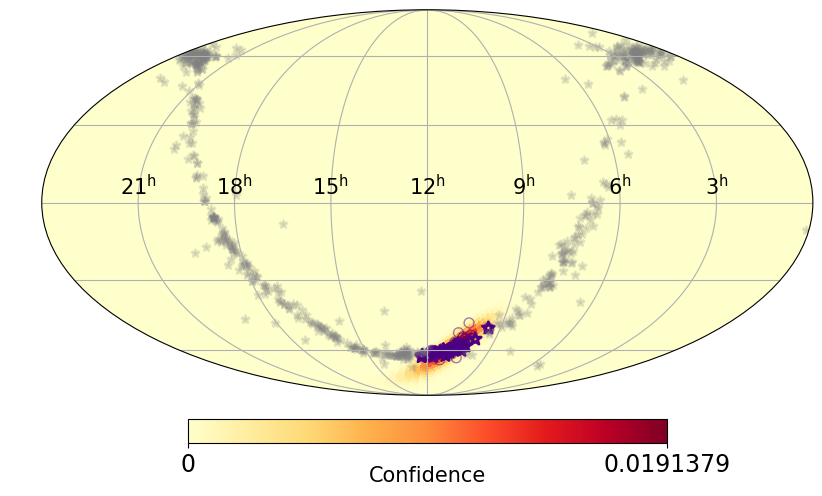}}\par 
   
   \subcaptionbox{Calculations based on X-46 (113$^{\circ}$, -22$^{\circ}$ and 7.6 kpc) as the progenitor. 105 stars lie within the 90\% credible region, with 2 matching the combined distance estimation and angular resolution. \label{fig:3}}{\includegraphics[width=1\linewidth]{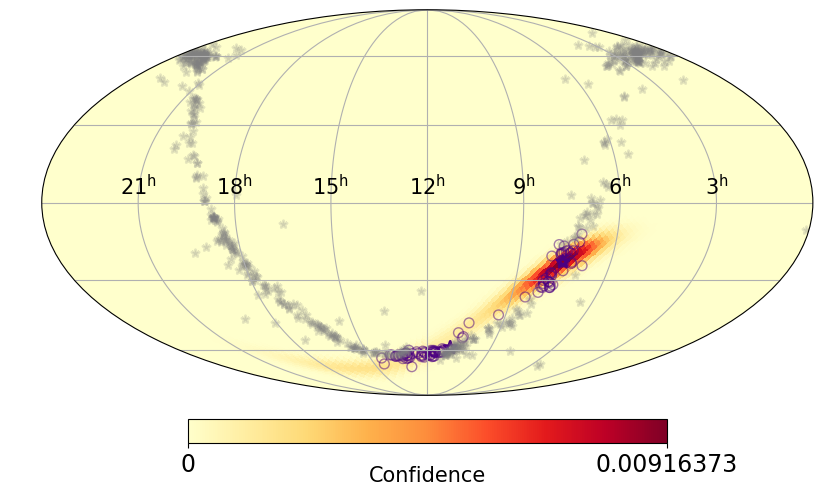}}\par
   
   \subcaptionbox{Calculations based on [W60] C10 (83$^{\circ}$, -67$^{\circ}$, 12.9 kpc) as the progenitor. 187 stars lie within the 90\% credible region, with 3 of those matching the combined distance estimation and angular resolution.\label{fig:4}}{\includegraphics[width=1\linewidth]{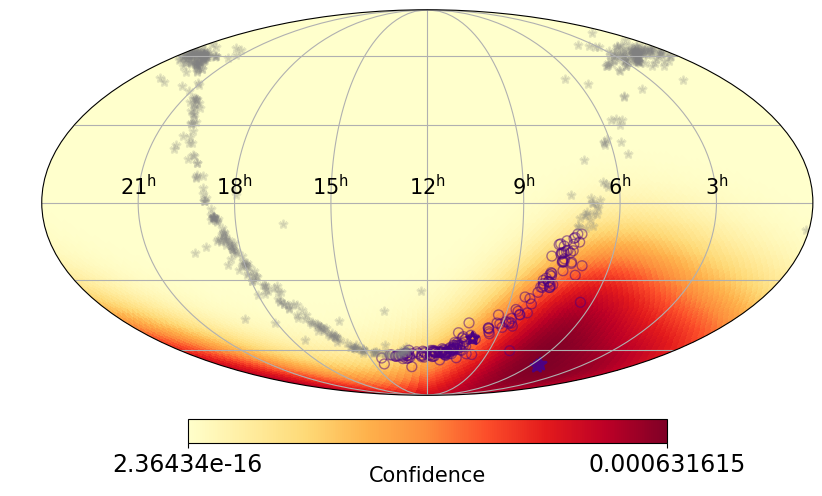}}\par
    \end{multicols}
    \caption{Confidence level localisation skymaps derived from triangulation, using the simulation of four CCSN events corresponding to the four stars shown in Fig.~\ref{distance_nu}. Overlaid is our RSG list (light gray stars), with those inside the 90\% angular error as dark purple open circles. RSGs, which also match the distance estimated from neutrinos, are shown by dark purple stars. The four panels show different sources at various RA, Dec, and distances. \label{fig:localization_maps}}
\end{figure*}

Neutrinos will be prolifically produced during stellar CC, and since they arrive minutes to days before the rise of the electromagnetic emission, provide the desired early alert \citep[see, e.g.,][]{Kharusi_2021}. Also, neutrino production rapidly rises during the last stages of nuclear burning, which can provide an advanced warning of the CC itself. One early established neutrino alert system is the SN Early Warning System \citep{Antonioli_2004}. SNEWS provides an early warning for a Galactic CCSN by tracking coincidences between neutrino experiments. In addition to providing coincident alerts, the successor to SNEWS, SNEWS 2.0, will aim at providing as much useful information as possible to capture the multi-messenger counterparts of the CC \citep{Kharusi_2021}. In particular, the distance and position of the source will be key parameters that can be estimated with the neutrino data. The {\tt snewpdag}\footnote{\url{https://github.com/SNEWS2/snewpdag}} software will be in charge of these calculations and used to release automated information in case of a CCSN alert.

SNEWS also coordinates with citizen astronomers, for instance, with the American Association of Variable Star Observers (AAVSO) to monitor potential CCSNe candidates long before they collapse, allowing more detailed observations on the development of stars as they evolve towards CC. The alert follow-up is being facilitated with the Recommender Engine For Intelligent Transient Tracking \citep{sravan_2020}. This is particularly useful as peculiar behavior prior to CC has been documented in several SNe with spurious observations. For example, the progenitor of SN 2009ip underwent a series of mini explosions prior to SN \citep{smith_2022}. It could also allow us to check for evidence of pre-SN mass loss, more accurately relate explosion energy with progenitors' initial mass, and define the relationship of light curve plateau brightness, temperature evolution of SN emission, and progenitor radius.

\subsection{Directional information from neutrinos} \label{pointing}

There are two main methods to obtain angular pointing information with the neutrino data: a direct direction reconstruction using anisotropic interactions and triangulation. The first exploits an interaction between the neutrino and a target with some intrinsic directionality with a detector that can exploit such information. For example, the electron scatters preferentially along the neutrino direction in neutrino-electron elastic scattering. Cherenkov detectors can track the direction through the light that is released when the electron is kicked. With this method, Super-Kamiokande can give the direction within 3--8 degrees for a SN at 10 kpc \citep{Mukhopadhyay_2020}. In relation to our catalog, neutrino pointing can reduce the possible potential stellar candidates on average within the error circles of 3-8 degrees down to 12-41 stars, respectively. Other neutrino interactions either do not retain very well directional information or have smaller cross sections. 

The second method is triangulation (or, more precisely, multilateration), which uses the relative arrival times of the neutrino burst through multiple detectors in different geographical locations to determine from where those neutrinos originate \citep{Vogel:1999kn_traingluation,Beacom_1999,Muhlbeier:2013gwa_traingulation}. The time of flight through the Earth for a neutrino is on the order of milliseconds. The success of the triangulation approach, therefore, depends on detectors registering samples of neutrino events large enough to achieve sufficient burst timing precision, or to have a very low background so as to precisely tell which is the first signal event. It also benefits from wide geographical spread among the contributing detectors. Estimates for the precision of triangulation under various assumptions and as a function of distance and location have been studied in various forms (e.g., \cite{linzer_2019,Kato:2020hlc_percisionoftraingulation,Hansen:2019giq_percisionoftraingulation,Coleiro:2020vyj_percisionoftraingulation}) where the most favorable conditions give a $1\sigma$ area of a few percent of the sky.

Once a trigger is received, SNEWS’s network runs the information through a pointing algorithm, producing a trigger packet of a skymap of direction probabilities and distance estimations (See \S\ref{DIST}). Direction information from anisotropic interactions, calculated by
individual experiments, can be incorporated into a final probability skymap,
which can then be used to prioritize follow-up observation, especially if one is able to
focus on a finite list of high-probability candidates, maximizing the prospects of obtaining both pre-SN observations and capturing the early rise of the SN with better sensitivity and cadence.

In Fig.~\ref{fig:localization_maps}, we show simulations of the triangulation error box estimated with the {\tt snewpdag} package and overlaid with our RSG list. We considered four stars ([W60] C10, X-46, CD-34 11794, and HD 303250) in the catalog covering different positions on the sky, distances, and stellar clustering. We fix the CC time epoch to illustrate both sky location dependence and the distance dependence. Simulations using SNEWPY (\cite{snewpy}) of the CCSN models in~\cite{Burrows:2020qrp},
with the corresponding stellar masses, have been used for this study. We apply the method described in \cite{Coleiro:2020vyj_percisionoftraingulation} to inverse beta decay ($\overline{\nu}_ep\rightarrow e^+n$) samples
at four detectors (Super-Kamiokande, JUNO, IceCube, and SNO+) to estimate the time delays. This method to estimate the time delay uncertainties by combining the observed neutrino lightcurves was chosen as it does not rely on any model template. The time uncertainties obtained range from 1ms, for distances up to $\sim$1kpc, to $\sim$10 ms for 12-13 kpc. We see that in all cases, the angular resolution significantly reduces the number of possible candidates, and further reduction can be achieved with the use of a distance estimate.

\subsection{Distance estimate with neutrinos} \label{DIST}

\begin{figure*}
    \centering
    \begin{multicols}{2}
   {\includegraphics[width=1\linewidth, height=5.95cm]{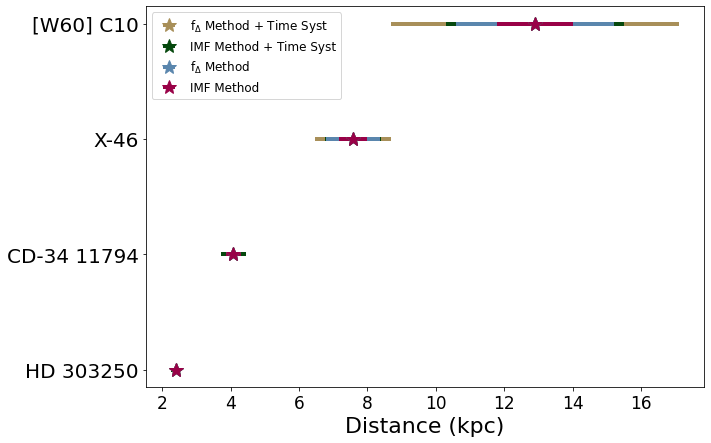}}\par 
   {\includegraphics[width=1\linewidth]{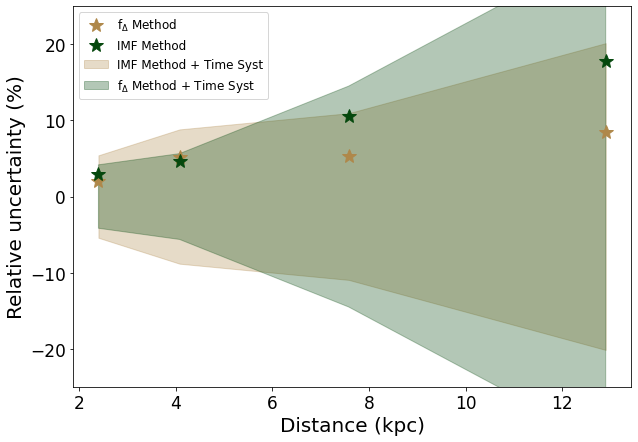}}\par 
    \end{multicols}
    \caption{On the right, we see the absolute distance uncertainties in kpc for our four selected stars, and for each of the two distance estimation methods. On the left, the relative distance uncertainty as a function of the distance. Star markers show the method (statistics plus systematics) error for each of the stars. The color bands correspond to the uncertainties when adding the systematics related to the timing uncertainty. \label{distance_nu}}
\end{figure*}

It was shown in \cite{Segerlund:2021dfz} that the $\overline{\nu}_e$ event rate in the early stages of the neutrino burst is related to the CCSN distance. Two methods proposed have been implemented into {\tt snewpdag}: the first compares the observed number of events ($n_{50}$) to the expected value weighted over the initial mass function (the ``IMF method''), and the second (the ``f$_{\Delta}$ method'') relies on a distance independent parameter, f$_{\Delta}$ = N(50)/N(100-150), and its linear relation with $n_{50}$ \citep{Horiuchi:2017qlw}. Here, N(50)/N(100-150) is the ratio of the expected number of events in the two respective time windows. We use both approaches to evaluate the distance uncertainty for our selected four candidate stars of the catalog, using a detector of the size of Super-Kamiokande or JUNO. We consider the same CCSN models as in \S\ref{pointing} and include the systematic uncertainty due to the model dependency of the methods. A further systematic uncertainty arises from the time uncertainty
of \S\ref{pointing}, which affects how the time windows (and therefore their event counts)
are reckoned. This systematic effect has been studied by shifting the neutrino light curve in simulations with respect to the true bounce time according to the expected time uncertainty. The results, shown in Fig.~\ref{distance_nu}, show the complementarity between the two methods: the f$_\Delta$ method, dominated by statistical uncertainty, works better
at smaller distances, while the IMF method outperforms at larger distances, albeit dominated by systematic uncertainties. Using this information on top of the triangulation skymap, the number of RSGs is further reduced in the best of cases to a few stars. For areas of high stellar density, the reduction is to less than 33\% of the candidates falling within the 90\% confidence level sky region, as seen in Fig.~\ref{fig:localization_maps}.

\subsection{Implications for MMA}

Obtaining detailed observations of a CCSN, from pre-explosion and neutrino burst through to shock breakout and the SN, would provide a treasure trove of data. Our RSG list would help realize this in the next Galactic CCSN by providing likely targets in the event of a neutrino alert in the future.

To date, pre-explosion observations of CCSN have been few and far between, but they have revealed interesting, unexpected behaviors. For example, the progenitor of SN2009ip underwent a series of mini explosions prior to SN \citep{maza_2009,miller_2009,Smith_2010} and SN1987a had a blue supergiant progenitor \citep{sonneborn_1987} contrary to the standard expectation for a RSG progenitor. Pre-explosion images have also revealed the so-called ``red supergiant problem'' wherein Type IIP SN progenitors appear to have a maximum mass of $\sim 17 M_\odot$, which is smaller than the most massive RSGs seen locally \citep{Smartt_2009}. Various possible solutions have been suggested. For example, the most massive RSGs may not be exploding as Type IIP SNe, necessitating a revision of our understanding of how RSGs end their lives, with deep implications for the formation rate of black holes and the influence of SN feedback in galaxy evolution \citep[e.g.][]{Horiuchi_2011,Crain_2015,Kochanek_2015}. Alternatively, the discrepancy could derive from our lack of understanding of stellar physics such as mass-loss \citep{Ekstrom_2012,Georgy_2012_3}, or larger than assumed systematic effects, for example, in bolometric corrections \citep{Davies_2013_5} or circumstellar extinction \citep{Walmswell_2012_6,Beasor_2016_7}, causing an underestimation of the progenitors mass. Having detailed pre-explosion monitoring data, in both cadence and multiple bands, would help in all the above debates.


As discussed above, the detection of GWs, multi-wavelength photons, and neutrinos provides a unique opportunity to probe the mechanisms of CC; however, the current setup of telescopes and detectors have limitations. Most Galactic SNe should be observable in the optical and virtually all should be observable in the near-IR  \citep{Adams_2013}. There is even a one-in-three chance for the SN to be visible to the naked eye. Neutrino detections easily cover the entire Milky Way galaxy. The detection by Super-Kamiokande with Gadolinium of a CCSN within 8.5 kpc from the Sun will have a pointing accuracy of $\sim 3$ degrees or better. Note that the region spanning a radius of 8.5 kpc includes 99\% of our full catalog. The detectability of GWs depends strongly on the uncertain core rotation. \citealt{Nakamura_2016} has shown that for an initially non-rotating core, GW detection is possible for sources up to the Galactic center if there is a coordinated observation with GW detectors and neutrino telescopes.



Statistically, the most likely distance to the next CCSN is close to the Galactic Center \citep{Adams_2013}. For example, the modeling of the progenitor and Galactic dust as a double-exponential spatial distribution indicates that within 5.8 kpc of the Galactic center corresponds to 62 percent of the Galactic CCSNe rate \citep{Adams_2013,Nakamura_2016}. 
While estimates from Galactic distributions of SN remnants (SNR) \citep{ahlers_2009,Mirizzi_2006} show some variation, a radius of $\sim$6 kpc consistently contains $\geq50\%$. In our sample, this results in 41 candidates. On the other hand, CCSN closer to the Sun will have better multi-messenger prospects, as we summarized in the previous paragraph. Here, we have 594 stars within 8.5 kpc from the sun (excluding the stars added based on uncertainty), i.e., within the distance regime where neutrinos can provide $\sim 3$ degrees of pointing accuracy and detection prospects for GWs and infrared photons are the most optimistic. Thus, the most promising objects are those within both regions; all 41 stars of the former set are also in the latter. Assuming that only M-type stars are approaching explosion following from the stellar evolutionary paths used in Section \ref{evol tracks}, 36 highly promising stars are retained within the overlapping spatial regions.


Further studies are guaranteed. Within the sample of 36 promising candidates, greater than $\rm >70\%$ have observations in more than 3 electromagnetic bands. Nevertheless, effort must be made to expand the available high-quality images to more wavelengths and maintain up-to-date follow-up spectra. 

\section{Summary and Future Work} \label{summary}
Using two complementary methods, one involving a thorough literature survey and the other using Gaia data, we compiled a catalog of 578 Milky Way RSGs. This is the largest catalog of its kind in the literature. To differentiate between RSGs and non-RSG contaminants, the most problematically extreme AGBs, we determined the bolometric luminosity and $\rm T_{eff}$. We compared them to stellar evolutionary tracks and Galactic AGB surveys. In this context, dust extinction is a major systematic uncertainty, which we estimated using {\tt mwdust}, a 3D dust map, to estimate dust extinction for each star to reduce and prevent correlations in uncertainties. The accuracy of this method was tested by comparing both the estimated extinctions and final luminosities to several alternative methods. Consideration of uncertainties adds another 62 stars that enter the RSG parameter space when individual errors are considered, resulting in a total of 640 candidates. 

Along with the information collected to determine their luminosities, the catalog contains details of each star's known characteristics, including evidence of binary interactions, magnetic fields, variability, and cluster membership. Using the RSG catalog compiled, we explored the RSG's spatial distribution, stellar radii range, and mass-loss rates. Finally, we explored the impacts of the RSG catalog for multi-messenger observations of the next Galactic CCSN. The intense neutrino burst from the CC of a massive star provides a natural alert, but our RSG catalog can help suggest locations for observations of a CCSN. Typically, a dozen or so of the RSGs reside within the typical pointing uncertainty given by the neutrino alert. 

Future work includes both improvement and expansion of the current catalog. As extinction is likely our greatest source of error, it seems clear that a better understanding of the complexity of Galactic ISM, including the completion of Galactic 3D dust maps and the excess of CSM around RSGs, such as a more robustly defined $\rm R_V$ and associated extinction ratios, would not only improve luminosity estimation but increase the likelihood of including dusty RSGs. Achieving the collection of more candidates in Region C is also important for future work. Issues of AGB contamination could be further mitigated with improvements in infrared spectra and flux values, like those that could be provided by the Roman Space Telescope scheduled to launch in 2027. The subsequent infrared CMDs can be combined with knowledge of an object's variability to reduce the ambiguity between AGBs and RSGs. Inspecting the new infrared CMDs can also provide an avenue for finding previously unknown candidates for further study. Expansion of available wavelength and SED fits will allow us to build a more robust profile for each of our candidate RSGs.

While we wait to detect neutrinos, the plan is to improve the characterization of candidates in our list so that post-explosion analysis will require as few assumptions and be as complete as possible. Along with our coordinated efforts to photometric variability monitoring via AAVSO, which is done primarily in the optical band, it would be important to expand the number of pre-explosion images across a greater range of wavelengths for each candidate. Mid-infrared photometry, in particular, will be useful to determine the dusty CSM around the RSGs, for which detailed knowledge of the environment will improve future analyses. This can be accomplished with the Roman infrared space telescope and its 2.4-meter mirror and Wide Field Instrument. Also important would be starting a campaign of follow-up spectroscopic observations to determine the variability, confirm our estimated luminosities, and collect spectra of the stellar progenitors as close to the explosion as possible. These efforts will also aid in determining fundamental properties like surface gravity and chemical composition to enhance our ability to trace stars from the zero-age main sequence to core collapse.


Some next steps are more immediate; for example, while a RSG is statistically the most likely next CC progenitor, other types of massive stars must also cause CCSNe, such as stripped-envelope SNe. This motivates the inclusion of a broader set of stars to complete further the target list \citep[e.g.,][]{2015MNRAS.447.2322R}. Continuation or initiation of monitoring more broadly CC candidates will provide more insight into each.

The next Galactic CCSN will be a once-in-a-generation opportunity to collect exquisite multi-messenger observations. With a more complete pre-assembled target list, the completeness of such observations will be improved and would likely shed new insights and potential surprises into stellar and CCSN physics.

\section*{Acknowledgements}
S.~Healy is supported by National Science Foundation (NSF) Grant No.~PHY-1914409 and No.~PHY-2209420. The work of S.~Horiuchi is supported by the U.S.~Department of Energy Office of Science under award number DE-SC0020262, NSF Grants AST-1908960, PHY-1914409 and PHY-2209420, and JSPS KAKENHI Grant Number JP22K03630 and JP23H0489, and
the Julian Schwinger Foundation. This work was supported by World Premier International Research Center Initiative (WPI Initiative), MEXT, Japan. 

The work by M. Colomer is supported by the F.R.S.-FNRS (Fonds de la Recherche Scientifique) through the research project IISN 4.4501.17 (40008230).
This work is also supported by the National Science Foundation “Windows on the Universe: the Era of Multi-Messenger Astrophysics” Program: “WoU-MMA: Collaborative Research: A Next-Generation SuperNova Early Warning System for Multimessenger Astronomy” through Grant Nos. 1914448, 1914409, 1914447, 1914418, 1914410, 1914416, and 1914426.

D.~M.\ acknowledges NSF support from grants PHY-1914448, PHY-2209451, AST-2037297, and AST-2206532. J.~T.\ is supported by the Science and Technology Facilities Council (STFC), UK.

We acknowledge with thanks the variable star observations from the AAVSO International Database contributed by observers worldwide and used in this research. We want to give special thanks to George Silvis whose efforts to produce AAVSO {\tt TargetTool} allow for more efficient monitoring of RSG candidates.

This work has made use of data from the European Space Agency (ESA) mission
{\it Gaia} (\url{https://www.cosmos.esa.int/gaia}), processed by the {\it Gaia}
Data Processing and Analysis Consortium (DPAC,
\url{https://www.cosmos.esa.int/web/gaia/dpac/consortium}). Funding for the DPAC
has been provided by national institutions, in particular, the institutions
participating in the {\it Gaia} Multilateral Agreement.

This publication makes use of data products from the Two Micron All Sky Survey, which is a joint project of the University of Massachusetts and the Infrared Processing and Analysis Center/California Institute of Technology, funded by the National Aeronautics and Space Administration and the National Science Foundation.

\section*{Data Availability}
Data for Table \ref{tab:sampleone} and Appendix A Table \ref{tab:finalone} is publicly available at \url{https://github.com/SNEWS2/candidate_list}.


\bibliographystyle{mnras}
\bibliography{rsg_mnras} 

\appendix

\setlength{\tabcolsep}{3pt}
\begin{landscape}
 \begin{table}
  \caption{Summary of calculated parameters for the highly probable and probable Galactic RSGs derived from either the Compliation-based or Gaia-based Method. Flags showing if an object has been identified as a Binary or Variable star are shown, but the full details are included in the unabridged catalog. The full table is publicly available at \url{https://github.com/SNEWS2/candidate_list}.
  \label{tab:finalone}}
  \begin{tabular}{lccccccccr}
 \hline
    Alias & SpType\_a &  $\rm log(\frac{L}{L_\odot})$ &  $\rm log(\frac{R}{R_\odot})$ &  log($\dot{M}$) & Binary\_Flag & Binary\_Ref\_\#\tnote{1} & Variable\_Flag &  Variable\_Ref\_\#\tnote{1} & Cluster/\\
    & & & & $log(M_\odot yr^{-1})$& & & & &Association \\
    \hline
        V* V348 Vel &M2 & 5.1 & 2.9 &-4.5 & & & Variable &  II, I, IV, III &      \\
        HD 237025 &M2 & 4.6 & 2.7 &-4.9 & Binary &III, IV & Variable &IV, III &Perseus Arm \\
        2MASS J18412383-0526073 &M0 & 4.2 & 2.4 &-5.6 & &  & Variable &    III & Stephenson 2 \\\relax
        [O66] 8:1003 &K5 & 3.2 & 1.9 &-6.6 & &  & &  &   \\
        UCAC2   5411346 &M1 & 3.9 & 2.3 &-5.8 & &  &  Variable &IV, III &   \\
        V* V778 Cas &M2 & 4.4 & 2.6 &-5.2 & &  & Variable & II, IV, III &    PER OB1 \\
        V* V466 Cas & M1.5 & 4.4 & 2.6 &-5.2 & &  & Variable &IV, III &    NGC 457 \\
        2MASS J19192791+1453516 &  M1.5 & 3.9 & 2.3 &-5.8 & Binary & III & Variable &IV, III &  \\
        V* V441 Per &M2 & 4.7 & 2.7 &-4.9 & &  & Variable & II, IV, III &    PER OB1 \\
        IRC +60091 &M2 & 4.7 & 2.7 &-4.9 & &  & Variable &  II, I, IV, III &    PER OB1 \\
        CD-61  3575 &M2 & 4.9 & 2.9 &-4.6 & Binary & V, 0 & Variable & II, IV, III &   \\
        IRAS 18104-1755 &K2 & 3.8 & 2.2 &-6.1 & &  &  Variable & II, IV, III &  Cluster near Hess J1813-178 \\
        BD+60   299 &M2 & 4.9 & 2.9 &-4.6 & &  & Variable &  II, I, IV, III &Perseus Arm \\
        DO 42086 &M4 & 4.8 & 2.8 &-4.7 & Binary & III & Variable &  II, I, IV, III &Perseus Arm \\
        IRAS 22096+5619 &M2 & 4.4 & 2.6 &-5.3 & Binary & III & Variable & III &Perseus Arm \\
        IRAS 01046+6309 &K3 & 4.3 & 2.4 &-5.6 & &  & &  &Perseus Arm \\
        V* AS Cep &M3 & 4.9 & 2.9 &-4.6 & &  & Variable &  II, I, IV, III &   \\
        DO 24315 &M3 & 4.4 & 2.6 &-5.2 &  Binary & II & Variable &  II, I, IV, III &Perseus Arm \\
        HD 174797 &     M2.5 & 4.1 & 2.4 &-5.5 & &  & Variable & II, IV, III &      \\
        HD 101779 &M0 & 3.9 & 2.3 &-5.8 & &  & Variable &IV, III &   \\\relax
        [2018MZM] 92 &     K5.5 & 4.2 & 2.4 &-5.6 & &  & Variable & II, I &   \\
        V* GS Vel &M2 & 4.0 & 2.4 &-5.6 & &  & Variable &IV, III &      \\
        HD  89736 &M0 & 4.0 & 2.3 &-5.8 & &  &  Variable &IV, III &      \\
        BM   VI   2 &M0 & 4.1 & 2.4 &-5.6 & &  & Variable &IV, III &   \\
        IRC +40427 &M1 & 4.3 & 2.5 &-5.4 & &  & Variable & II, IV &      \\
        IRC -30312 &     M2.5 & 5.4 & 3.1 &-4.2 & &  & Variable &  II, I &      \\
        UCAC2 09087349 &  M2.5 & 4.1 & 2.4 &-5.5 & &  & Variable &IV, III &   \\
        IRAS 22483+5713 &K3 & 4.0 & 2.3 &-5.8 & Binary & III & Variable &    III &Perseus Arm \\
        V* IM Cas &M2 & 5.0 & 2.9 &-4.6 & Binary & III & Variable & II, IV, III &Perseus Arm \\
        HD  45829 &K1 & 3.9 & 2.2 &-6.0 & Binary & III & Variable &  III &      \\
        BD+10  3764 &M3 & 4.2 & 2.5 &-5.4 & Binary & III & Variable & II, IV, III & CAS OB7 \\\relax
        [A72c]  37 &M0 & 4.5 & 2.6 &-5.2 & &  & &  &   \\
        V* EH Cas &M3 & 4.9 & 2.8 &-4.7 & Binary & III & Variable &     II, IV, III &Perseus Arm \\
        V* AZ Cas &K5 & 4.2 & 2.5 &-5.5 & Binary &  VI, V, 0, ... &    Variable &IV, III &  CAS OB8 \\
        V* V605 Cas &M2 & 4.7 & 2.8 &-4.8 & &  & Variable & II, IV, III &    PER OB1 \\
        HD 237006 &M1 & 4.4 & 2.6 &-5.3 & Binary & V, 0, 36, ... &  Variable &IV, III &    PER OB1 \\
        IRAS 22285+5706 &M2 & 4.2 & 2.5 &-5.4 & Binary & III & Variable &IV, III &Perseus Arm \\
        IRAS 00375+6400 &  M1.5 & 4.2 & 2.5 &-5.5 & Binary &  III & Variable &II, III &Perseus Arm \\
        CD-24 13583 &M3 & 4.2 & 2.5 &-5.4 & &  & Variable &IV, III &   \\
        UCAC2 05561199 &K7 & 4.9 & 2.8 &-4.8 & &  & Variable &IV, III &   \\
        BD+62   281 &M3 & 4.7 & 2.8 &-4.8 & &  & Variable & II, IV, III &Perseus Arm \\
        V* AV Per &M2 & 3.8 & 2.3 &-5.8 & Binary & III & Variable &  II, I, IV, III &  CEP OB1 \\
        HD 163755 &M2 & 4.3 & 2.5 &-5.3 &  Binary & III & Variable &IV, III &      \\
    \hline
  \end{tabular}
    \begin{tablenotes}\footnotesize
        \item[1]  I \cite{Holl_2018_gaia_dr2_variability}, II \cite{Eyer_2022_gaia_dr3_variability,halbwachs_2022,gaia_non_single_Stars_2022}, III AAVSO, IV \cite{shappee_2014,jayasinghe_2020}, V \cite{panta_2020}, VI \protect\cite{Neugent_2019}, VII \cite{gvaramadze_10.1093/mnras/stt1943}, VIII \cite{grunhunt_2010}, IX \cite{tessore_2017}. Those denoted by numbers are listed in Table \ref{tab:sampleone}.
    \end{tablenotes}
 \end{table}
\end{landscape}

\section{Sample Characteristics}\label{appendixList}

Table \ref{tab:finalone} lists 44 stars of our final sample, showing select columns. The full table is available in machine-readable form on Vizier and at \url{https://github.com/SNEWS2/candidate_list}. The columns of the full table are:

Column (1): Alias

Column (2): Gaia DR3 source id 

Column (3): Gaia DR2 source id

Column (4): Adopted mean spectral type

Column (5): Spectral type(s) 

Column (6): Spectral type references (format is only the last name of first author and year)

Columns (7)-(8): Either the same as reference if the spectral type was derived in that reference or the source cited for the spectral type given in the reference and associated number used in tables

Columns (9)-(10): Effective temperature and associated errors based on spectral types 

Columns (11) - (23): Values taken from 2MASS [detailed column descriptions can be found at the link: \href{https://www.ipac.caltech.edu/2mass/releases/allsky/doc/sec2_2a.html}{Cal Tech user guide}] 

Columns (24)-(58): Values taken from Gaia DR3 [detailed column descriptions can be found at the link: \href{https://gea.esac.esa.int/archive/documentation/GDR3/Gaia_archive/chap_datamodel/sec_dm_main_source_catalogue/ssec_dm_gaia_source.html}{Gaia DR3 documentation}]

Columns (59)-(61): Values related to Gaia EDR3 distance taken from \cite{bailer_jones_2021}

Columns (62)-(85): Values taken from Gaia DR2 [detailed column descriptions can be found at the links: \href{https://gea.esac.esa.int/archive/documentation/GDR2/Gaia_archive/chap_datamodel/sec_dm_main_tables/ssec_dm_gaia_source.html}{Gaia DR2 documentation} and \href{https://gea.esac.esa.int/archive/documentation/GDR2/Gaia_archive/chap_datamodel/sec_dm_main_tables/ssec_dm_ruwe.html}{Gaia DR2 RUWE documentation}]

Columns (86)-(89): Values related to Gaia DR2 distance taken from \cite{Bailer_Jones_2018} 

Columns (90)-(91): Extinction in the K band and associated error 

Column (92): Extinction in the H band 

Column (93):  Extinction in the J band 

Columns (94)-(95): Bolometric correction for $K_{s}$ band and associated error 

Columns (96)-(98): Distance module and associated upper and lower errors 

Column (99): Extinction in the V band 

Columns (100)-(101): Intrinsic $K_{s}$ mag and associated error 

Column (102): Intrinsic H mag 

Column (103): Intrinsic J mag

Columns (104)-(106): Absolute K magnitude and associated upper and lower errors 

Columns (107)-(109): Bolometric magnitude and associated upper and lower errors 

Column (110): Stellar radius in log form

Column (111): Stellar radius in units of solar radii

Column (112): Log of luminosity in terms of solar luminosity 

Column (113): Log of effective temperature 

Columns (114)-(116): Galactic coordinates in kiloparsecs 

Columns (117): Mass-loss based on Eq. \ref{mass loss equation}

Columns (118)-(130): Variable values taken from Gaia DR3 [detailed column descriptions can be found at the link: \href{https://gea.esac.esa.int/archive/documentation/GDR3/Gaia_archive/chap_datamodel/sec_dm_variability_tables/}{Gaia DR3 variable documentation}

Columns (131)-(140): Non-single star values taken from Gaia DR3 [detailed column descriptions can be found at the link: \href{https://gea.esac.esa.int/archive/documentation/GDR3/Gaia_archive/chap_datamodel/sec_dm_non--single_stars_tables/ssec_dm_nss_two_body_orbit.html}{Gaia DR3 non-single star documentation}

Columns (141)-(147): Variable values taken from Gaia DR2 [detailed column descriptions can be found at the link: \href{https://gea.esac.esa.int/archive/documentation/GDR2/Gaia_archive/chap_datamodel/sec_dm_variability_tables/}{Gaia DR2 variable documentation}

Columns (148)-(153): Variable values taken from AAVSO detailed column descriptions can be found at the link: \href{https://vizier.cds.unistra.fr/viz-bin/VizieR-3?-source=B/vsx&-out.max=50&-out.form=HTML\%20Table&-out.add=_r&-out.add=_RAJ,_DEJ&-sort=_r&-oc.form=sexa}{AAVSO VSX Catalog}

Columns (154)-(159): Variable values taken from ASAS-SN V detailed column descriptions can be found at the link: \href{http://asas-sn.osu.edu/variables}{ASAS-SN Variable Star Database}

Columns (160)-(163): Variable flag, details on a star's variable status, reference, and each reference's symbol in Table \ref{tab:finalone}

Columns (164)-(167): Binary flag, details on a star's binary status, reference, and each reference's symbol in Table \ref{tab:finalone}

Columns (168)-(170): Measured magnetic field flag, reference, and each reference's symbol in Table \ref{tab:finalone}

Columns (171)-(173): Runaway flag, reference, and each reference's symbol in Table \ref{tab:finalone}

Column (174): Cluster or OB association

Column (175): Method for collection (Compilation-based, Gaia-Based, within error bars, or Close star addendum)

Column (176): Simbad Classification 

Column (177): Region (A-E) \label{regioncolumn}

\section{References for Close star Table}\label{list of close star ref}
The references used to create the table detailing all stars whose errors in Gaia are too large to retain or are not included, but are generally accepted as RSGs throughout the literature include: \citealt{Levesque_2005_17_ref_15,tessore_2017,gaia_dr3_summary_2022,grunhunt_2010,vlemmings_2005,bidelman_1951,keenan_1989_ref_26,Messineo_2019,leeuwen_2007,tremko_2010,lockwood_and_wing_1982,pesch_1967,apell_2021,hohle_2010,Levesque_massey_2020,cox_2012,harper_2008,shinnaga_2017,stock_2018,mathias_2018,wittkowski2012,tabernero_2021,avvakumova_2013,lopez_2018,tetzlaff_2011,moravveji_2013,Carpenter_1999,joyce_2020,harper_2001,pecaut_2012}, and \citealt{daviesandbeasor_2020}.

\bsp	
\label{lastpage}
\end{document}